\documentclass[preprint2]{aastex}
\usepackage{graphicx}
\usepackage{lscape}
\usepackage{txfonts}

\shorttitle{Dynamics of rotation in M dwarfs}
\shortauthors{Houdebine \& Mullan}

\begin{document}

\title{Dynamics of rotation in M dwarfs: Indications for a change in the 
dynamo regime in stars at the onset of complete convection\thanks{Based on 
observations available at Observatoire de Haute Provence and the European 
Southern Observatory databases and on Hipparcos parallax measurements.}}

\author{E.R. Houdebine\altaffilmark{1}}
\affil{Armagh Observatory, College Hill, BT61 9DG Armagh, Northern Ireland}

\and 
\author{D.J. Mullan\altaffilmark{2}}
\affil{Department of Physics and Astronomy, University of Delaware, Newark, 
DE 19716, USA}

%DRAFT: 4 Février 2015
 
\begin{abstract}
We have measured $v\sin i$ with high precision for a sample of dM3 stars (86 
targets). We detected rotation in 82 stars (73 dM3 stars, and 9 dM3e stars). 
We compare our measurements of $v\sin i$ for all the stars in our dM0, dM2, 
dM3 and dM4 samples to those from other authors. We find a good agreement 
down to $v\sin i$ values of less than 1 $km ~s^{-1}$. The mean of the 
differences between measurements is only 0.42~$km~ s^{-1}$.

We find that the distribution of $P$/sin$i$ for our dM3 stars is different 
from the distribution of $P$/sin$i$ among our samples of dM2 and dM4 stars. 
The mean rotation rate for the dM3 stars (excluding dM3e and sdM3 stars) is 
significantly slower (25.8 days) than for dM2 (14.4 days) and dM4 stars (11.4 
days). Analogous behavior also emerges among the faster rotators (dMe stars): 
we find that a longer rotation period also occurs at spectral sub-type dM3e. 
Our data suggest that, as regards the rotational properties of lower main 
sequence stars, spectral sub-type dM3 stands out as exhibiting unusual slow 
rotation compared to that of adjoining sub-types. Our data lead us to suggest 
that the unusual rotational properties of M3 dwarfs may represent a signature 
of the transition to complete convection (TTCC).
\end{abstract}


\section{Introduction}

There is a close correlation between stellar rotational period $P$ and 
magnetic activity levels in cool dwarfs over a certain range of rotation 
periods (Soderblom 1982, Vogt et al. 1983, Marcy \& Chen 1992, Patten \& Simon 
1996, Fekel 1997, Delfosse et al. 1998, Jeffries et al. 2000, Pizzolato et al. 
2003, Mohanty \& Basri 2003). The sense of this "rotation-activity 
correlation" (RAC) is as follows: the slower a star rotates (i.e. the longer 
its $P$), the weaker is the level of magnetic activity in that star (other 
things being equal). Rather than using the period $P$ alone, it has been found 
that the Rossby number $R_{0} \sim P/\tau_{c}$ (where $\tau_{c}$ is the 
convective overturning time) is a useful parameter to quantify the 
rotation-activity correlation in the case of the X-ray luminosity and the 
Ca\,{\sc ii} line core fluxes (e.g. Noyes et al. 1984). The reason why $R_0$ 
is particularly useful in quantifying the correlation is that value of $R_0$ 
is thought to parameterize how the magnetic dynamo depends on rotation. 
Although a more recent study by Reiners et al. (2014) suggests that 
$L_{X}/L_{bol}$ correlates better with $P^{-2}R^{-4}$ (where $R$ is the 
stellar radius) than with $R_0$, we find it convenient in the present paper to 
deal with $R_0$: the main reason for our choice is that recent efforts to 
model a dynamo in a deep spherical shell (such as the convective envelopes
which exist in M dwarfs) have reported their results as a function of $R_0$ 
(Gastine et al. 2012). We shall refer to some of the results from these dynamo 
models below (Section 7.2).

The well-defined RAC mentioned in the above papers extends only over a finite 
range of periods: when $P$ becomes shorter than a certain value $P_{c}$ (of 
order several days), the level of activity no longer increases as rapidly with 
decreasing $P$. Instead, for $P<P_{c}$, the activity level tends towards 
"saturation" (e.g. Walter 1982; Vilhu and Rucinski 1983). In terms of $R_0$, 
saturation is found to set in for numerical values of $R_0$< 0.1 (Reiners et 
al 2009). The onset of saturation at short periods may be associated with 
nonlinear effects of some kind, or with the existence of a finite upper level 
of mechanical energy in the convection zone (Mullan 1984), or with a 
saturation in the field strength which is generated by the dynamo mechanism 
(Reiners et al 2009; Yadav et al. 2013). Whatever the physical reason(s) for 
the saturation, it seems likely that, with the reduced sensitivity of activity 
to rotational period in the saturated regime, the properties of dynamo 
operation might not be as easy to extract and identify with confidence in that 
regime (where non-linear effects may dominate) as those properties are in the 
unsaturated region (where linear behavior may dominate). In the latter region, 
with $P$ values longer than a few days, activity levels decrease rapidly as 
$P$ becomes longer (varying as $P^{-4}$ [Walter 1982] in the case of K 
dwarfs): in these "unsaturated" conditions, the data may contain more readily 
accessible information about the dynamo properties, and about the processes 
which lead to rotational braking. For this reason, we focus most of our 
attention in this paper on "unsaturated" {\it slowly rotating} M dwarfs, i.e. 
those with rotational periods which are {\it longer} than $P_{c}$. As a result 
of this selection, most of the stars in our sample have low levels of 
activity.    

Only a few previous studies have included slowly rotating field stars (e.g. 
Marilli et al. 1986, Reiners 2007,  Kiraga \& Stepien 2007). Houdebine and 
collaborators (this paper, Papers XV, XVIII, Houdebine et al. 2015) have 
attempted to measure $v\sin i$ in slow rotators for a few narrow spectral 
domains; dK5, dM0, dM2, dM3 and dM4. They successfully measured $v\sin i$ for 
a number of stars at these spectral types and derived empirical RAC's. 

The principal goal in this paper is the empirical determination of the 
rotational properties of low mass stars in a range of spectral types extending 
from dK5 to dM4: within this range, our goal is to present data for certain 
narrow ranges of sub-types, including dK5, dM0, dM2, dM3, and dM4. A secondary 
goal is to present a qualitative analysis of our results in terms of various 
physical processes which contribute to the loss of angular momentum from stars.
One of the physical processes which we consider has to do with the linear size 
of magnetic loops on M dwarfs, and how these lengths change particularly 
between M2 and M4 (see Section 7.4).

\subsection{Stellar rotation: empirical}

In this paper, we report on an empirical rotational distribution which 
we have obtained for a sample of 86 field M dwarfs. The stars in our sample 
have been chosen specifically to be confined within a narrowly delimited range 
of spectral sub-types: all of the target stars for which we present data in 
the present paper are classified as spectral sub-type M3. Our principal goal 
is to compare and contrast the rotational properties at M3 with those of 
neighbouring sub-types M2 and M4: in previous papers, rotational data have 
been published for those neighbouring sub-types.
  
An independent and extensive study of rotational distributions, including 
quantitative theoretical modelling of angular momentum loss, has been reported 
by Reiners and Mohanty (2012: hereafter RM). Stars in the RM study span a much 
broader range of spectral types than we study here: the RM sample extends from 
solar type (G2) down to late M. The RM sample also included stars in clusters 
(with ages up to 650 Myr) as well as field stars (with ages estimated to be on 
average 3 Gyr). In the present study, by confining our attention to field 
stars only, we expect that our sample stars are on average older than the 
oldest cluster reported by RM, i.e. older than 650 Myr.  

\subsection{Stellar rotation: theory} 

We first make general comments on the loss of angular momentum from a 
star. Then we summarize the extensive and detailed modelling efforts reported 
by Reiners and Mohanty (2012: hereafter RM).

\subsubsection{General} 

The linear speed v$_{r}$ (or angular velocity $\Omega$) with which a star 
rotates at any particular time in its history is determined by the interplay 
of a number of factors. (i) The initial conditions which accompanied the 
formation of the star: the faster the proto-stellar cloud was rotating, the 
faster the star is likely to be rotating today. (ii) Any variations in the 
moment of inertia of the star $I$ (which controls the angular momentum 
$J = I\Omega$) in the course of evolution will cause reciprocal variations in 
$\Omega$ in the absence of torques. (iii) If finite torques are in fact at 
work (e.g. due to mass loss), then the instantaneous value of $dJ/dt$ is 
non-zero. The integrated effects of $dJ/dt \neq$0 during the entire course of 
previous evolutionary time will alter the value of $J$, and therefore also 
the value of $\Omega$ at any instant of time.  

An example of how $\Omega$ in a particular stellar model (with 
mass=$1M_{\odot}$) varies with time as a result of particular specifications 
of the various factors can be seen in MacDonald and Mullan (2003:  
hereafter MM03) (their Fig.~2): from an initial rotation period of 16 days, 
the model at first speeds up to a rotation period as short as about 1 day 
at an age of several 10’s of Myr , i.e. just as the 1 solar mass star 
reaches the main sequence. This pre-main-sequence speed-up in rotation 
is associated with a reduction in I as the star 
contracts. Subsequently, during main sequence evolution, the period slows 
down, reaching periods of 20-30 days at ages of several Gyr. The slowing down 
on the main sequence is associated with angular momentum loss due to the 
coupling between a stellar wind and the star's global magnetic field. The 
latter probably arises from the operation of a dynamo of some kind inside the 
star.

Further examples of how $\Omega$ varies with time for stars of various 
masses can be found in RM  (see their Figure 2). One of the RM examples refers 
to the same mass (1 solar mass) as that modelled by MM03, except that RM chose 
an initial rotation period of 8 days rather than the 16 days of MM03.  The 
major features reported by MM03 are clearly present in the RM results: there 
is an initial speed-up in the rotation, which reaches its shortest period at 
ages of 40-50 Myr, and then the period  becomes progressively longer, reaching 
periods of order 30 days at the present age of the Sun. For stars of lower 
mass, similar features occur, except that it takes longer for the lower mass 
stars to reach the main sequence: in the case of the 0.1 M$_{\odot}$ model, 
the shortest rotation does not occur until an age of 600-700 Myr, which is the 
time required for such a low mass star to reach the main sequence.

\subsubsection{The modeling work of Reiners and Mohanty (RM)}

RM have reported on a modelling study in which they make predictions of 
rotational periods for low mass stars at various stages of evolution. Their 
model is based on the following expression for the rate at which the angular 
momentum $J$ of a star rotating with angular velocity $\Omega$ varies with 
time due to mass loss: $dJ/dt$ = -(2/3) $\dot{M}$ $\Omega r_{A}^{2}$ . Here, 
$\dot{M}$ is the rate of mass loss, and $r_{A}$ is the Alfven radius, i.e. the 
radial location in the wind where the local wind speed $v$ equals the local 
Alfven speed $v_{A}$ . The distance $D$ out to which the magnetic field of the 
star is energetically capable of enforcing co-rotation on the wind is equal to 
$r_{A}$: the distance $D$ is therefore a measure of the "lever arm" over which 
the magnetic field can interact strongly enough with the wind to exert torques 
on the star, thereby braking the star's rotation. The larger $D$, the more 
effective the braking of the star becomes. The numerical value of $D$ =  
$r_{A}$ for any particular star depends on the field strength $B_{o}$ on the 
surface of the star: the latter depends in turn (via some kind of dynamo 
mechanism) on the angular velocity of the star. RM’s approach to modelling  
$B_{o}$ as a function of $\Omega$ includes two segments: these segments are 
meant to model the two segments of the RAC mentioned in Section 1 above, i.e. 
unsaturated at slow rotation, and saturated at fast rotation. To model this, 
RM assume that at slow rotations ($\Omega < \Omega_{crit}$ ), $B_o$ increases 
with increasing $\Omega$ according to a power law $B_o \sim \Omega^{a}$ (where 
a = 1-2), but at faster rotations ($\Omega > \Omega_{crit}$ ), $B_o$ is 
independent of $\Omega$, saturating at a constant value $B_{crit}$. (In terms 
of the discussion in Section 1 above, $P_{c}$ and $\Omega_{crit}$ are related 
via $P_{c}$ = 2$\pi$/$\Omega_{crit}$.) Adopting evolutionary tracks from the 
literature, and assuming that $v_{A}$ equals $K_{V} v_{e} (r)$ (i.e. $v_{A}$ 
is proportional to the gravitational escape speed $v_{e}$ from the star at 
$r_{A}$),  RM had most of the ingredients to  calculate the evolution of $J$ 
for a star of a given mass. For the remaining ingredients, RM  adopted  a 
radial field, and assumed that $\dot{M}$, $B_{crit}$, $\Omega_{crit}$, and 
$K_{V}$  take on the same numerical values for stars of all masses, and that 
these numerical values also remain constant with time.  RM found that the 
evolution of $J$ depends mainly on their choice of numerical values for two 
quantities: $\Omega_{crit}$ and a certain constant $C$: the latter depends on 
$B_{crit}$, $K_{V}$, and $\dot{M}$.

RM presented empirical data on rotational distributions as a function of both 
mass and spectral type for various star samples in which the ages were 
estimated by different methods. E.g. rotation periods of 0.1-10 days were 
shown for clusters with known evolutionary ages (5 Myr, 130 Myr, and 650 Myr). 
For older stars, i.e. field stars, the ages of young-disk stars (ages 1-5 Gyr) 
could be assigned with some confidence based on kinematics. However, the field 
sample also included some old-disk/halo stars, although RM argued that the 
presence of such old-disk stars should not skew the results significantly. The 
best-fit values $\Omega_{crit}$  and $C$ allowed RM to achieve significant 
success in replicating several “crucial” (the word used by RM) features of the 
rotational distributions of stars in clusters at ages of 130 Myr and 650 Myr, 
as well as for stars in the field. In addition, they were successful in 
replicating the result reported by West et al (2008) that activity life-times 
increased from $<$1 Gyr in the earliest M dwarfs to several Gyr in the latest 
M dwarfs.

\subsection{Reasons for studying the spectral range dM2-dM4}

In this paper, we focus on a much narrower range of spectral types than RM did.
The reason we pay particular attention to the spectral sub-types from dM2 
to dM4 has to do with the fact that models of M dwarfs on the main sequence 
are expected to exhibit a transition to complete convection (TTCC) at spectral 
sub-types M3-M4 (e.g. Mullan and MacDonald 2001). It is believed that stars 
with a radiative core (including the Sun) have dynamos which are dominated 
by processes at the interface between the core and the convective envelope. 
However, in a star which is completely convective, an interface dynamo 
cannot exist. Therefore, at the TTCC, the dynamo is expected to switch from 
interface operation to a different regime (e.g. a distributed dynamo). 
If a change in dynamo regime occurs at the TTCC, perhaps there is also an 
accompanying change in the angular momentum loss rate $dJ/dt$. If we can 
obtain rotational distributions which are sufficiently detailed, it may be 
possible to identify a signature of a change in $dJ/dt$. The key question in 
the present paper is: is it possible to narrow down this signature to a 
specific spectral sub-type?

\subsection{Summary of the present paper}

In the present paper, we do not plan to re-do the evolutionary modeling 
of RM. Our goal is more restricted: we wish to report on high-precision 
rotational velocities for a sample of narrowly delimited spectral subtypes. A 
secondary goal is to analyze the results, in a way which is less detailed than 
that used by RM.  

Our approach is actually consistent with some cautionary remarks made by RM 
themselves. RM admitted that the number of assumptions and parameters which 
enter into their formalism is large enough that there is still room for 
improvement in reaching a quantitative understanding of stellar rotation. They 
note that “Any discrepancies that arise in the comparison of our model to the 
data will motivate an examination of” the assumptions {\it a posteriori}. 
Among the items listed by RM where they consider that there is room for 
improvement are the following. (1) The assumption that stars of all masses 
have the same values of $\dot{M}$, $B_{crit}$, $\Omega_{crit}$, and $K_{V}$, 
and that these values remain unchanged at all evolutionary times. (2) The 
assumption of purely radial field geometry. 3) The assumption that $v_{A}$ 
scales linearly with $v_{e}$. (4) The numerical values of $v\sin i$ for M 
dwarfs are in many cases too small to be reliably measured: as a result, it is 
hard to determine from the data where exactly the transition from saturation 
to unsaturation occurs. (5) The magnitude of $\dot{M}$ in M dwarfs may be 
quite different from the solar value: RM cite a source which claims that M 
dwarf stars have $\dot{M}$ values which are perhaps orders of magnitude larger 
than the solar value. (6) In the RM theoretical formalism, the mass of the 
star is used as the primary parameter: however, at the latest spectral types, 
RM note that different methods of estimating the mass of an M dwarf may differ 
by up to 30\%.

In the present work, we use our own data to address some of these items. 
E.g., as regards item (2), we examine (in Section 7.3 and 7.4) the non-radial 
properties of fields in M dwarfs. As regards item (3), we examine (in Section 
7.1) a possible rationale for how the Alfven radius might vary from one M 
dwarf to another. As regards item (4), a principal goal of the present work 
(Section 4) is to report on $v\sin i$ values which push beyond the limits of 
detectability which had constrained some of the RM data, thereby providing a 
better chance to possibly detect the transition from saturated to unsaturated 
dynamo behaviour. As regards item (5), we consider (in Section 7.1) some 
evidence which suggests that $\dot{M}$ in M dwarfs may be smaller than the 
solar value. As regards item (6), we prefer to use colors and spectral types
(rather than masses) to identify which stars should be included in our sample: 
these are immediately accessible to observation (Section 2), and our data sets 
do not rely on knowing the mass of any star in the sample. As regards item (1),
 our approach can make no comment.

In this paper, we report on empirical rotational distributions which have been 
obtained for field stars at spectral sub-type M3. With field stars, we expect 
that our sample stars are significantly older than the oldest cluster reported 
by RM, i.e. older than 650 Myr.  The stars which are of most interest to us 
(at spectral type M3), arrive on the main sequence (MS) at ages of ~200 Myr 
(see RM). By confining attention to stars with ages older than 650 Myr, our 
sample stars would have reached the main sequence long ago. To the extent that 
this is valid, it is meaningful for us to consider testing the theoretical 
prediction that on the main sequence, stars become fully convective at spectral
 types in the vicinity of M3.

\section{Observations}

\subsection{Selection of our sample stars}

In previous studies (e.g. Houdebine \& Stempels 1997 Paper VI, Houdebine 2008 
Paper VII, Papers XV, XVIII) we have chosen to isolate samples of stars 
according to their effective temperatures, i.e. based on photospheric 
conditions. This photospheric emphasis was shown to be necessary even when we 
were interested in modelling chromospheres in M dwarfs: we found that the 
formation of chromospheric lines depends on the photospheric conditions as 
indicated by $T_{eff}$. At first sight, this may seem somewhat paradoxical, 
but in support of this claim, we offer the following. 

If we were studying specifically stars which had highly active chromospheres, 
then it would be true that the formation of chromospheric lines would be 
dominated by the way in which mechanical energy deposition controls local 
conditions in the chromosphere. In such stars, the Balmer lines are strongly 
in emission.  However, in choosing our sample of stars, we have {\it not} 
relied on selecting stars with high activity (see Section 2.2 below). In fact 
the opposite is true. The sample includes a significant number of stars which 
were included in the HARPS archives because they had been observed 
specifically in order to search for exoplanets. In such searches, the presence 
of a high level of magnetic activity is typically used to exclude stars from 
the exoplanet sample: variability associated with activity can obscure the 
variability associated with a putative planet motion. As a result, there could 
be a bias in our sample towards stars with low levels of activity. This has an 
important consequence in the context of an early M dwarf: in such a star, the 
photosphere generates Balmer lines which are weakly in absorption. But if a 
low level of mechanical flux is deposited in the atmosphere of such a star, 
models predicted a result which is somewhat surprising: the effect is not to 
generate Balmer lines in emission. On the contrary, the effect is to drive 
the Balmer lines more strongly into absorption (Cram \& Mullan 1979, 
Houdebine \& Stempels 1997). The equivalent width of H$_{\alpha}$ in 
absorption reaches a maximum value (about 0.7 \AA\ in an M0 star) at a certain 
flux of mechanical energy. Subsequent increases in mechanical flux result at 
first in filling in the absorption. Eventually, when the mechanical energy 
flux exceeds a critical value, the Balmer lines do indeed go into emission. 
These predictions were verified by Stauffer and Hartmann (1986) in their 
observations of about 200 M dwarfs in the Gliese catalog. Thus, in M dwarfs 
with low activity levels, such as many of the stars in the HARPS exoplanet 
archive, the Balmer spectrum retains the absorption characteristic which is 
most often associated with the photosphere. The study of the effect of the 
effective temperature on the H\,{\sc i} and Ca\,{\sc ii} spectral lines can 
also be found in Houdebine \& Doyle (1994) for spectral types from G to M.

As regards the current study of the rotation of M dwarfs, the photospheric 
conditions once again play an important role: indeed we showed in previous 
studies (Paper VII, Houdebine 2010b Paper XIV, Houdebine 2011b Paper XVI, 
Houdebine 2012 Paper XVII, Houdebine et al. 2015) as well as in the present 
study that the rotation properties of M dwarfs depend on the effective 
temperature (which correlates with radius and mass, Kiraga \& Stepien 2007). 
Therefore, it is crucial if one wants to investigate the rotation in M dwarfs 
to isolate stellar samples with the same effective temperature and not 
intermingle multiple spectral types such as in Nielsen et al. (2013). In order 
to isolate a sample of dM3 stars with the same effective temperature, we 
decided to select our stars according to their R-I colour. We showed in 
previous papers (cited above) that this colour is a good effective temperature 
indicator and has relatively low sensitivity to metallicity. Moreover R-I data 
are widely available for nearby stars and have a high precision.

All the stars in our selected sample have similar R-I colours (see Table 1), 
and the same effective temperature of 3290 K to within $\pm$60 K with only a 
few exceptions. Our choice of R-I colours corresponds approximately to the 
spectral type dM3 (Leggett 1992). We note, however, that spectral 
classification may differ from one author to another. In this regard, in 
previous Papers on dM1 stars (Paper VII, XII, Houdebine 2009b Paper XIII, 
Paper IX, Paper XIV, Houdebine 2010c Paper X, Paper XV, Houdebine et al. 2012 
Paper XIX), we used a different calibration. According to their infrared 
colours and the classification of Leggett (1992) these stars are in fact dM2 
stars. We shall use in the future this calibration and therefore all our 
previous work on "dM1" stars in this series of papers should be considered as 
papers referring to dM2 stars. 

We selected stars with (R-I)$_{c}$ (R-I in the Cousins system) in the range 
[1.284;1.416] with a few exceptions close to this range, which also 
corresponds to (R-I)$_{K}$ (R-I in the Kron system) in the range [1.003;1.113] 
according to the transformation formulae of Leggett (1992) (see Leggett 1992 
for more information on the Cousin's and Kron photometric systems). Values of 
(R-I) were taken from the following papers: Veeder (1974), Eggen (1974), 
Rodgers \& Eggen (1974), Eggen (1976a, 1976b), Eggen (1978), Eggen (1979), 
Eggen (1980), Weis \& Upgren (1982), Upgren \& Lu (1986), Eggen (1987), Booth 
et al. (1988), Leggett \& Hawkins (1988), Bessel (1990), Weis (1991a 1991b), 
Dawson \& Forbes (1992), Leggett (1992), Weis (1993). 

We selected a list of 381 dM3 stars from measurements of R-I in the 
literature. We searched the European Southern Observatory (ESO) and 
Observatoire de Haute Provence (OHP) databases for spectroscopic \'echelle 
observations of these stars. In these databases, we found observations for 
86 distinct dM3 stars from two different \'echelle spectrographs; HARPS 
(High Accuracy Radial velocity Planet Search, ESO) and  SOPHIE (OHP). 

\subsection{Biases in our sample stars}

The stars in our sample include all of the dM3 stars which are present in the 
databases from all observing programs with HARPS and SOPHIE. These programs 
are for the most part planet-search programs, but they also include some 
programs on magnetic activity. Our sample of stars is brightness limited: as 
such, there is a bias towards the brighter M dwarfs, i.e., towards stars 
within the spectral subtype M3 with closer distances and/or larger radii.

We find that our initial target list is complete to a visual magnitude of 
13.5. This corresponds to an absolute magnitude of 11.45 (the limit from 
normal dwarfs to subdwarfs) at a distance of 26 pc for main-sequence disk 
stars. For halo stars (M3 subdwarfs) the target list is complete only to a 
much smaller distance because of the intrinsic faintness of these objects. We 
find that among our target list, all dM3 stars of visual magnitude $\sim$12 or 
brighter have been observed and are in the present sample of stars. For an 
absolute magnitude of 11.45, this implies a maximum distance of 13 pc. 
Therefore, we believe that our stellar sample is complete for old disk and 
young disk stars out to a distance of about 13 pc.

As far as the bias towards bright stars is concerned, we found it has little 
effects on the mean rotation period of low activity dM3 stars (see Sect.~6). 
This is due to the fact that the rotation period does not change much with the 
stellar radius for disk stars (see Sect.~6). For the same reason, this bias, 
although important, has little consequences on the RAC. We emphasize here that 
our calculated mean rotation periods (reported below in Section 6) refer to 
disk stars only. We do not include M3 subdwarfs.

In the planet-search programs in the HARPS and SOPHIE databases, the observers 
tend to avoid very active stars because their spot modulation mimic the 
effects of planets and adds considerably to noise in the data. Therefore our 
stellar sample is biased towards low activity stars. However, we found that 
some 10% of our sample are dM3e stars: in the dM3e stars, the activity level 
is enhanced to some extent compared to the dM3 stars. The important point is 
that all M3 dwarfs brighter than m$_{v}$=12 seem to have been observed, and 
our sample is somewhat biased towards low activity stars.

As far as other spectral types are concerned (dK5, dM0, dM2 and dM4), we rely 
on the same properties of the rotation period as a function of radius 
(Houdebine 2010, Paper XIV, Houdebine 2011, Paper XVI, Houdebine 2012, Paper 
XVII). Therefore, for all these spectral types we do not expect the bright 
star bias to present a problem in determining the mean rotation period of slow 
rotators. Also, for all these spectral types except the dM0 spectral type, the 
sample includes up to about 10% of active stars: as a result, the 
rotation-activity relationships which we derive should be reliably 
inter-comparable among the various subtypes. We have only a slight problem 
with dM0 stars, because the sampling of active stars includes only one dM0(e) 
star and one dM0e star. Therefore the rotation-activity relationship 
is not as well defined at high activity levels for this spectral type (see 
Houdebine et al. 2015). Nevertheless, this has no significant effect on the 
conclusions of the present study.

We emphasize that for all stars in our subsamples from dM2 to dM4, we have 
used the same relationship between R-I color and effective temperature, and 
the same relationship between absolute magnitude and radius (see Section 3 
below). As a result, even if there may be uncertainties in the individual 
numerical values of ${\it T_{eff}}$ and radius which we derive from these 
relationships, nevertheless, in a ${\it differential}$ study involving a 
comparison between stars of various subtypes from dK5 to dM4, the relative 
uncertainties should be much smaller. It is precisely such a differential 
study that leads to the result which we regard as among the most significant 
in the present paper (see Fig.~10 below).

\subsection{Spectrographs and data reduction}

The two high-resolution \'echelle spectrographs whose data are used in the 
present paper were intended for planet search programs and were designed to be 
extremely stable in wavelength and resolution. This stability makes them well 
suited for the measurement of rotational velocities in slowly rotating M 
dwarfs. 

HARPS is a fiber-fed high resolution echelle spectrograph mounted on the 3.6m 
telescope at ESO-La Silla Observatory. HARPS is the ESO facility for the 
measurement of radial velocities with the highest precision currently 
available (about 0.001 $km s^{-1}$). In order to achieve this goal, this 
spectrograph is fed by a pair of fibers and is optimized for mechanical 
stability. It is contained in a vacuum vessel to avoid spectral drift due to 
air temperature and pressure variations. The resolving power of HARPS  is 
115,000. The spectral range covered is 378-691 nm. An automated reduction 
procedure reduces the spectra collected and yields a precise radial velocity. 
A more complete description of this spectrograph can be found in Mayor et al. 
(2003).

SOPHIE is a cross-dispersed echelle spectrograph located in a 
temperature-controlled room at OHP. The spectra cover the wavelength range 
387-694 nm. The instrument is computer controlled and a standard data 
reduction pipeline automatically reduces the data upon CCD readout. The 
spectrograph resolution is 75,000, and the precision in radial velocities is 
0.002-0.003 $km s^{-1}$ which makes it one of the most stable spectrographs in 
the world. A more complete description of this spectrograph can be found in 
Perruchot et al. (2008).

\section{Stellar parameters}

The methods we use to derive the effective temperature and the stellar radius 
for each star are the same as in our previous studies on M dwarfs.
In order to derive effective temperatures (T${eff}$) for dM3 stars, we used 
the calibration of Jones et al. (1994) for M dwarfs. We used the formulae:

\begin{equation}
T_{eff} = -799\times (R-I)_{c} + 4372 K
\end{equation}

\noindent
This relation is valid down to (R-I)$_{c}$=2.3. Jones et al. (1994) 
quote a precision of $\pm$50 K on their temperature determinations.

In order to derive the stellar radii, we used the classical formula;

\begin{equation}
M_{v}+BC = 42.36 - 5\times log\frac{R}{R_{\odot}} - 10\times log T_{eff},
\end{equation}

\noindent
where symbols take their usual meaning. The value of BC varies with stellar 
temperature. We used the BC tabulation with (R-I)$_{c}$ from Kenyon \& 
Hartmann (1995) in order to calculate BC. The stellar radii and the effective 
temperatures are listed in Table 1 for all our target stars.

% Table 1
%\begin{landscape}
\begin{table*}
\caption[ ]{Parameters of dM3 stars in our sample, including the $S/N$ ratio 
for the sum of the spectra for each star, as well as the number of spectra. The 
full table is available from the authors upon request.}

\begin{center}

\tiny
\begin{tabular}{|lccccccccccc|} \hline

Star    & v   &(R-I)$_{c}$&     Teff      &Spectral&     $\pi$     & Distance &       Mv     &    R$_{\star}$   &  No.of &  S/N  &  S/N   \\
        &(mag) & (mag)     &     (K)      & Type   &     (m'')     &   (pc)   &     (mag)    &   (R$_{\odot}$)  & spec.&       &        \\
        &      &           &              &        &               &          &              &                  &      & HARPS & SOPHIE \\
        &      &           &              &        &               &          &              &                  &      &@5000\AA&@5000\AA \\
GJ 1046 &11.595&   1.292   & 3350 $\pm$50 & dM3    & 71.06$\pm$3.23&  14.07   &10.85$\pm$0.09& 0.426$\pm$0.029  &  6   &  61   &    -   \\
GJ 1050 &11.730&   1.322   & 3320 $\pm$50 & dM3    & 66   $\pm$13  &  15.2    &10.83$\pm$0.39& 0.450$\pm$0.085  &  9   &  69   &    -   \\
GJ 1097 &11.456&   1.400   & 3250 $\pm$50 & dM3    & 81.38$\pm$2.49&  12.29   &11.01$\pm$0.06& 0.461$\pm$0.026  &  7   &  71   &    -   \\
GJ 1125 &11.710&   1.125   & 3470 $\pm$50 &sdM3    &103.46$\pm$3.94&   9.67   &11.78$\pm$0.09& 0.224$\pm$0.015  &8,5,3 &  68   &63HR,62HE\\
GJ 1203 &12.158&   1.352   & 3290 $\pm$50 & dM3    & 59.63$\pm$3.55&  16.77   &11.04$\pm$0.12& 0.425$\pm$0.035  &  8   &  44   &    -   \\
GJ 1212A&12.032&   1.403   & 3250 $\pm$50 & dM3    & 53.41$\pm$4.76&  18.72   &11.42$\pm$0.19& 0.383$\pm$0.043  &  1   &  10   &    -   \\
GJ 1212B&12.032&   1.403   & 3250 $\pm$50 & dM3    & 53.41$\pm$4.76&  18.72   &11.42$\pm$0.19& 0.383$\pm$0.043  &  1   &  10   &    -   \\
GJ 1271 &11.704&   1.298   & 3340 $\pm$50 & dM3    & 47.52$\pm$2.82&  21.04   &10.09$\pm$0.12& 0.612$\pm$0.049  & 4,4  &   -   &61HR,79HE\\
GJ 2121 &12.280&   1.386   & 3270 $\pm$50 & dM3    & 44.58$\pm$5.36&  22.43   &10.53$\pm$0.24& 0.563$\pm$0.074  &  1   &   8   &    -   \\
GJ 3139 &11.770&   1.333   & 3310 $\pm$50 & dM3    & 50.68$\pm$2.53&  19.73   &10.29$\pm$0.11& 0.585$\pm$0.045  &  2   &  29   &    -   \\
GJ 3160A&12.01&    1.292   & 3340 $\pm$50 & dM3    & 38.45$\pm$3.15&  26.01   &10.69$\pm$0.18& 0.461$\pm$0.049  &  1   &  10   &    -   \\
GJ 3160B&12.01&    1.292   & 3340 $\pm$50 & dM3    & 38.45$\pm$3.15&  26.01   &10.69$\pm$0.18& 0.461$\pm$0.049  &  1   &  10   &    -   \\
GJ 3189 &12.67 &   1.318   & 3320 $\pm$50 &sdM3    & 95.5 $\pm$10.9&  10.5    &12.57$\pm$0.23& 0.201$\pm$0.026  &  9   &  42   &    -   \\
GJ 3279 &11.8  &   1.324   & 3310 $\pm$50 & dM3    & 65   $\pm$13  &  15      &10.86$\pm$0.40& 0.445$\pm$0.086  & 10   &  69   &    -   \\
GJ 3293 &12.0  &   1.308   & 3330 $\pm$50 & dM3    & 55   $\pm$9   &  18      &10.70$\pm$0.33& 0.469$\pm$0.078  & 19   &  89   &    -   \\
GJ 3404A&11.73 &   1.412   & 3250 $\pm$50 & dM3    & 64   $\pm$9   &  16      &10.76$\pm$0.29& 0.525$\pm$0.079  & 19   & 102   &    -   \\
GJ 3412A&11.028&   1.393   & 3260 $\pm$50 &sdM3    & 95.43$\pm$2.36&  10.48   &11.68$\pm$0.05& 0.335$\pm$0.017  &  1   &   -   &  41HR  \\
GJ 3412B&11.028&   1.393   & 3260 $\pm$50 &sdM3    & 95.43$\pm$2.36&  10.48   &11.68$\pm$0.05& 0.335$\pm$0.017  &  1   &   -   &  41HR  \\
GJ 3459 &11.712&   1.390   & 3260 $\pm$50 &sdM3    & 94.31$\pm$3.31&  10.60   &11.58$\pm$0.08& 0.349$\pm$0.022  &  7   &  63   &    -   \\
GJ 3528 &11.728&   1.300   & 3330 $\pm$50 & dM3    & 51.25$\pm$3.72&  19.51   &10.28$\pm$0.15& 0.563$\pm$0.053  &  6   &  59   &    -   \\
GJ 3563 &11.953&   1.311   & 3330 $\pm$50 & dM3    & 63.47$\pm$3.54&  15.76   &10.97$\pm$0.11& 0.415$\pm$0.032  & 11   &  66   &    -   \\
GJ 3598 &12.46 &   1.292   & 3340 $\pm$50 & dM3    & 45   $\pm$9   &  22      &10.73$\pm$0.39& 0.452$\pm$0.085  &  1   &   7   &    -   \\
GJ 3634 &11.95 &   1.291   & 3340 $\pm$50 & dM3    & 56   $\pm$11  &  18      &10.69$\pm$0.39& 0.460$\pm$0.087  & 49   & 145   &    -   \\
GJ 3643 &12.369&   1.346   & 3300 $\pm$50 & dM3    & 50.90$\pm$4.58&  19.65   &10.90$\pm$0.19& 0.450$\pm$0.050  &  1   &   8   &    -   \\
GJ 3708A&11.709&   1.376   & 3270 $\pm$50 & dM3    & 79.43$\pm$2.36&  12.59   &11.21$\pm$0.06& 0.406$\pm$0.023  &  8   &  63   &    -   \\
GJ 3846 &12.257&   1.399   & 3250 $\pm$50 &sdM3    & 70.03$\pm$4.89&  14.28   &11.48$\pm$0.15& 0.371$\pm$0.036  &  6   &  44   &    -   \\
GJ 3892 &11.473&   1.340   & 3300 $\pm$50 & dM3    & 69.19$\pm$2.60&  14.45   &10.67$\pm$0.08& 0.496$\pm$0.032  &  8   &  74   &    -   \\
GJ 3916A&11.279&   1.306   & 3330 $\pm$50 & dM3    & 66.21$\pm$3.18&  15.10   &11.13$\pm$0.11& 0.383$\pm$0.029  &  3   &  42   &    -   \\
GJ 3916B&11.279&   1.306   & 3330 $\pm$50 & dM3    & 66.21$\pm$3.18&  15.10   &11.13$\pm$0.11& 0.383$\pm$0.029  &  3   &  42   &    -   \\
GJ 4004 &12.1  &   1.455   & 3210 $\pm$50 &sdM3    & 81   $\pm$16  &  12      &11.64$\pm$0.39& 0.373$\pm$0.071  &  2   &  28   &    -   \\
GJ 4129 &11.95 &   1.304   & 3330 $\pm$50 & dM3    & 63.1 $\pm$3.8 &  15.8    &10.95$\pm$0.13& 0.416$\pm$0.036  &  1   &  11   &    -   \\
Gl 12   &12.61 &   1.415   & 3240 $\pm$50 &sdM3    & 84   $\pm$11  &  12      &12.23$\pm$0.27& 0.268$\pm$0.038  & 6,2  &  37   &  37HE  \\
Gl 70   &10.915&   1.274   & 3350 $\pm$50 & dM3    & 87.62$\pm$2.00&  11.41   &10.63$\pm$0.05& 0.463$\pm$0.024  & 7,7  &  98   & 105HR  \\
        &      &           &              &        &               &          &              &                  &      &       &        \\

\hline 

\end{tabular}
\end{center}
\end{table*}
%\end{landscape}

\normalsize

\small
% Table 2
\small
\begin{center}
\begin{table*}
  \caption[ ]{Results for $v\sin i$ and $P$/sin$i$ for M3 stars, including 
standard deviations. We put in parenthesis the detections below $3\sigma$. 
Also listed: FWHMs of the cross-correlation line profiles. The 
full table is available from the authors upon request.}
\begin{tabular}{|lcccccc|}

 \hline

Star    &     FWHM      &       FWHM   &$v\sin i\pm\sigma$&$v\sin i\pm\sigma$&Mean $v\sin i$& $P/\sin i\pm\sigma$ \\
        &  (\AA$\pm$)   &  (\AA$\pm$)   & (km s$^{-1}$)    &    (km s$^{-1}$) &(km s$^{-1}$)&  (days)    \\
        &     HARPS     &    SOPHIE     &      HARPS       &      SOPHIE      &        &                 \\
GJ 1046 & 0.1273 0.0011 &     -         &  2.63$\pm$0.064  &        -         &  2.63  & 8.20$\pm$0.76   \\
GJ 1050 & 0.1063 0.0011 &     -         & (0.59+0.26-0.52) &        -         &  0.59  &(38.4-17+40)     \\
GJ 1097 & 0.10505 0.0011&     -         &      0.00        &        -         & $<$0.5 &$>$47            \\ 
GJ 1125 & 0.1063 0.0011 & 0.1401 0.0014 & (0.59+0.26-0.52) &  1.39+0.20-0.24  &  0.99  &(11.5-1.5+2.4)    \\ 
GJ 1203 & 0.1072 0.0011 &     -         & (0.82+0.29-0.18) &        -         &  0.82  &(26.2-6.5+18)    \\
GJ 1212A& 0.1066 0.0011 &     -         & (0.69+0.22-0.44) &        -         &  0.69  &(28.1-9.2+40)    \\
GJ 1212B& 0.1077 0.0011 &     -         & (0.90+0.16-0.22) &        -         &  0.90  &(22.0-5.2+9.9)   \\
GJ 1271 &      -        & 0.1401 0.0014 &         -        &  1.39+0.20-0.24  &  1.39  & 22.3-4.4+6.8    \\
GJ 2121 & 0.1187 0.0011 &     -         &  2.06+0.08-0.18  &        -         &  2.06  & 13.9-2.3+3.3    \\
GJ 3139 & 0.1077 0.0011 &     -         & (0.90+0.16-0.22) &        -         &  0.90  &(32.8-7.0+14)    \\
GJ 3160A& 0.1080 0.0011 &     -         &  0.95+0.15-0.18  &        -         &  0.95  & 24.5-5.5+9.0    \\
GJ 3160B& 0.1088 0.0011 &     -         &  1.06+0.14-0.16  &        -         &  1.06  & 22.0-4.6+6.6    \\
GJ 3189 & 0.1125 0.0011 &     -         &  1.42$\pm$0.09   &        -         &  1.42  & 7.18$\pm$1.2    \\ 
GJ 3279 & 0.1074 0.0011 &     -         & (0.85+0.17-0.26) &        -         &  0.85  &(26.4-8.6+19)   \\
GJ 3293 & 0.1074 0.0011 &     -         & (0.85+0.17-0.26) &        -         &  0.85  &(27.9-8.5+19)   \\
GJ 3404A& 0.1077 0.0011 &     -         & (0.90+0.16-0.22) &        -         &  0.90  &(29.5-8.2+15)   \\
GJ 3412A&      -        & 0.135  0.0034 &         -        &      $<$0.5      & $<$0.5 &        -        \\ 
GJ 3412B&      -        & 0.134  0.0034 &         -        &      $<$0.5      & $<$0.5 &        -        \\ 
GJ 3459 & 0.1066 0.0011 &     -         & (0.69+0.22-0.44) &        -         &  0.69  &(25.7-7.4+40)   \\ 
GJ 3528 & 0.1077 0.0011 &     -         & (0.90+0.16-0.22) &        -         &  0.90  &(31.7-7.2+14)   \\
GJ 3563 & 0.1074 0.0011 &     -         & (0.85+0.17-0.26) &        -         &  0.85  &(24.7-5.7+13)   \\
GJ 3598 & 0.1066 0.0011 &     -         & (0.69+0.22-0.44) &        -         &  0.69  &(33.2-13+40)    \\
GJ 3634 & 0.1074 0.0011 &     -         & (0.85+0.17-0.26) &        -         &  0.85  &(27.3-8.8+19)   \\
GJ 3643 & 0.1083 0.0011 &     -         &  1.00+0.14-0.18  &        -         &  1.00  & 22.8-5.0+8.1    \\
GJ 3708A& 0.1060 0.0011 &     -         & (0.50+0.30-0.50) &        -         &  0.50  &(41.1-17+40)    \\ 
GJ 3846 & 0.1072 0.0011 &     -         & (0.82+0.18-0.29) &        -         &  0.82  &(22.9-5.9+16)   \\ 
GJ 3892 & 0.1072 0.0011 &     -         & (0.82+0.18-0.29) &        -         &  0.82  &(30.7-7.1+19)   \\
GJ 3916A& 0.1102 0.0011 &     -         &  1.24$\pm$0.14   &        -         &  1.24  & 15.6-2.6+3.3    \\ 
GJ 3916B& 0.1447 0.0011 &     -         &  3.51$\pm$0.05   &        -         &  3.51  & 5.52$\pm$0.50   \\ 
GJ 4004 & 0.1063 0.0011 &     -         & (0.59+0.26-0.52) &        -         &  0.59  &(32.0-14+40)    \\ 
GJ 4129 & 0.1060 0.0011 &     -         & (0.50+0.30-0.50) &        -         &  0.50  &(42.0-18+40)    \\
Gl 12   & 0.1080 0.0011 &     -         &  0.95+0.15-0.18  &        -         &  0.95  & 14.2-3.6+5.9    \\ 
Gl 70   & 0.1074 0.0011 & 0.1388 0.0014 & (0.85+0.17-0.26) & (1.17+0.24-0.26) &  1.01  &(23.2-4.3+7.0)   \\ 
Gl 109  &      -        & 0.1398 0.0014 &        -         &  1.34+0.22-0.24  &  1.34  & 15.6-2.8+4.3    \\ 
Gl 145  & 0.1074 0.0011 &     -         & (0.85+0.17-0.26) &        -         &  0.85  &(22.0-4.6+11)   \\ 
Gl 163  & 0.1074 0.0011 &     -         & (0.85+0.17-0.26) &        -         &  0.85  &(29.1-6.2+15)   \\ 
Gl 204.2& 0.1066 0.0011 &     -         & (0.69+0.22-0.44) &        -         &  0.69  &(40.2-13+40)    \\ 
Gl 207.1& 0.3141 0.0011 & 0.3318 0.0014 &  9.27$\pm$0.034  &  9.77$\pm$0.05   &  9.52  & 2.81$\pm$0.25   \\
Gl 226  &      -        & 0.1385 0.0014 &         -        & (1.12+0.24-0.27) &  1.12  &(21.8-4.6+8.3)   \\ 
Gl 238  & 0.1074 0.0011 &     -         & (0.85+0.17-0.26) &        -         &  0.85  &(31.4-6.7+16)   \\
Gl 251  &     -         & 0.1386 0.0014 &        -         & (1.14+0.24-0.27) &  1.14  &(18.9-3.9+6.8)   \\ 
GL 298  & 0.1083 0.0011 &     -         &  1.00+0.14-0.18  &        -         &  1.00  & 27.4-5.0+8.3    \\
GL 352  & 0.1527 0.0011 &     -         &  3.89$\pm$0.05   &        -         &  3.89  & 8.51$\pm$0.65   \\ 
Gl 357  & 0.1077 0.0011 &     -         & (0.90+0.16-0.22) &        -         &  0.90  &(20.8-3.9+7.8)   \\ 
Gl 358  & 0.1100 0.0011 &     -         &  1.21$\pm$0.14   &        -         &  1.21  & 19.8-2.9+3.6   \\ 
        &               &               &                  &                  &        &                 \\
\hline 
\end{tabular}
\end{table*}
\end{center}
\normalsize

\clearpage

\section{Determination of the projected rotation periods}

% Fig. 1
\begin{figure*} 
\vspace{-1.5cm}
\begin{centering}
\hspace{-4.5cm}
\includegraphics[width=16cm,angle=-90]{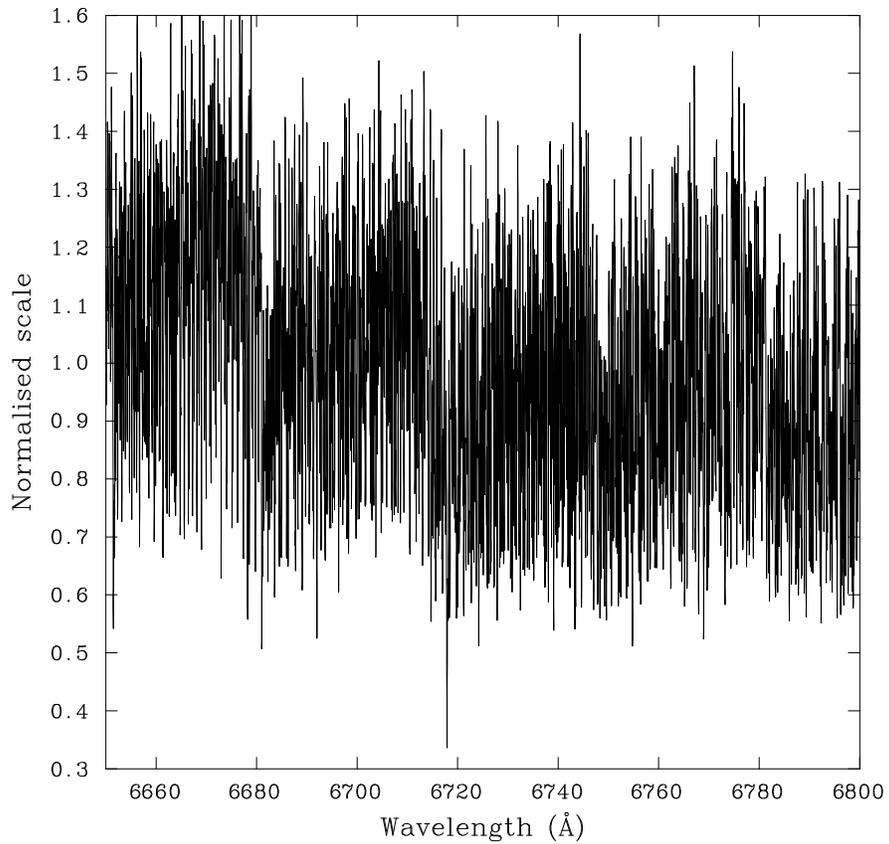}
\vspace{-0.5cm}
\end{centering}
\caption[]{The spectral range 6650\AA\ - 6800\AA\ used for our 
cross-correlations for the star Gl 436: this range yields very clean 
cross-correlation profiles at sub-type dM3. The $S/N$ ratio for Gl 436 is 
about 800 in this range. The variations relative to the continuum are not 
noise: they are due to numerous blended weak absorption lines. The mean flux 
in this spectrum has been normalised to 1 for clarity.}
\end{figure*}

% Fig. 2
\begin{figure*} 
\vspace{-1.5cm}
\begin{centering}
\includegraphics[width=12cm,angle=-90]{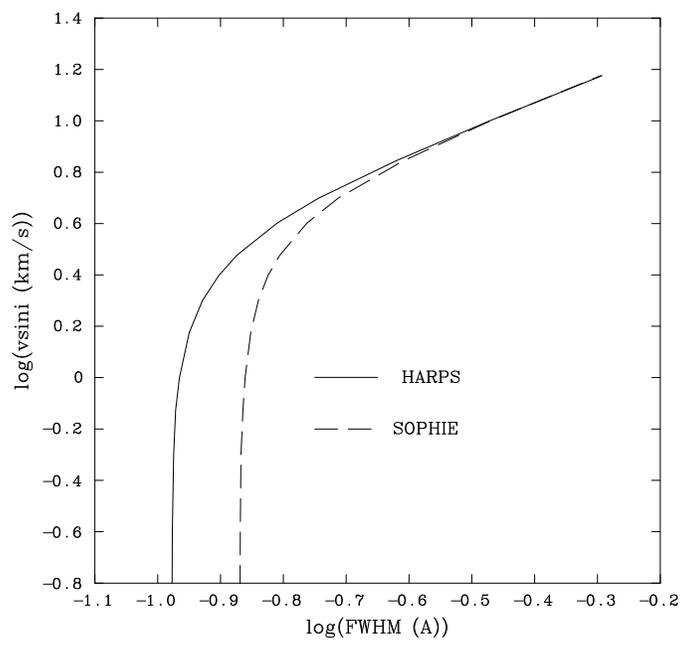}
\end{centering}
\vspace{-0.5cm}
\caption[]{Plots of the conversion between the measured FWHM (in \AA) of the 
cross-correlation peaks and the numerical value of $v\sin i$. Solid line: 
HARPS data. Dashed line: SOPHIE data.}
\end{figure*}

In this paper, we use the same technique as in previous studies in order to 
determine the projected rotational broadening $v\sin i$: we use a 
cross-correlation technique for a selected narrow spectral range. The choice 
of the spectral range is crucial in order to obtain cross-correlation profiles 
that are as clean as possible, as well as having low background noise (see 
also Sect.~5). For dM3 stars, we used a 150\AA\ wide spectral domain in the 
red end of the spectrum, between 6650\AA\ and 6800\AA\ (see Fig. 1, which 
shows the spectrum of one particular star in our dM3 sample: Gl 436). In this 
wavelength range, the presence of many weak and unsaturated narrow spectral 
lines yield clean cross-correlation functions as well as a low background 
noise. We have found that the spectral lines in the red yield the best 
cross-correlation results for dM2, dM3 and dM4 stars. 

Because our stars all have closely similar spectral types, it is possible to 
intercompare the spectra and look for possible rotational broadening 
effects with the highest precision. The broadening of the lines depend on 
three main parameters; rotational broadening, and micro- and macro-
turbulence. The latter parameters (probably associated in some way with the 
turbulent motions of convectively rising/falling gas, and with numerical 
values of typically 1 km/s) are not expected to vary much among our narrowly 
selected sample of dM3 stars. Metallicity plays a role in the strengths of the 
lines, but less significantly in their shapes. Therefore, changes in spectral 
line widths from one star to another are expected to depend mainly on 
differences in rotational velocity from star to star.

The procedure of selecting appropriate templates for low $v\sin i$ values is 
the same as in previous papers. We found that the best template for HARPS is 
GJ 1097 with FWHM of only 0.10505\AA\ (4.69 km/s). This is broader than the 
0.1006\AA\ FWHM (5.46 km/s) we measured for dM2 stars. However this increased 
broadening is consistent with the different wavelength domains we used in 
calculating the cross-correlations (for dM2 stars we used the wavelength range 
5460\AA\ - 5585\AA ). For SOPHIE, we found that the best template is Gl 655 
with a FWHM of the cross-correlation peak of 0.1349\AA .

In order to measure $v\sin i$, we cross-correlated the stellar spectra with 
the template spectrum. We then subtracted the cross-correlation background 
following the same method as in Paper XIV. We then measured the FWHM of the 
cross-correlation peaks. In order to recover $v\sin i$ from the FWHM we then 
proceeded as follows: We computed the theoretical rotational profiles 
including limb darkening for the following series of $v\sin i$ values (in 
units of $km\ s^{-1}$): 0.00, 0.025, 0.25, 0.50, 0.75, 1.00, 1.50, 2.00, 2.50, 
3.00, 4.0, 5.0, 7.0, 10.0 and 15.0. We then convolved the theoretical 
rotational profiles with the cross-correlation profile of GJ 1097 for HARPS 
and Gl 655 for SOPHIE. We measured the FWHM of the convolved theoretical 
rotational profiles and show these measurements as a function of $v\sin i$ in 
Fig.~2 for HARPS and SOPHIE. We use this diagram to derive $v\sin i$ values 
from our calculated FWHM result for each star. The measured FWHM, $v\sin i$, 
and $P$/sin$i$ values are listed in Table~2. The values in parenthesis are 
possible values but with a large uncertainty (see next Section).

We found that non-zero values for $v \sin i$ could be obtained for most of our 
dM3 stars; the values of $v\sin i$ were found to be typically in the range 
0.5-2 $km\ s^{-1}$. The $v\sin i$ values are smaller on average for dM3 stars 
than for dM2 and dM4 stars. We find that on average, $v\sin i$=1.05 $km\ 
s^{-1}$ for dM3 stars (excluding spectroscopic binaries, sdM3 and dM3e stars), 
whereas we found $v\sin i$=2.41 $km\ s^{-1}$ for dM2 stars and $v\sin i$=1.57 
$km\ s^{-1}$ for dM4 stars. As we shall see below (Fig. 6), this also implies 
longer rotation periods for dM3 stars than for dM2 and dM4 stars. 

With HARPS, we also observed a few stars of particular interest; six 
spectroscopic binaries (GJ 1212AB, GJ 3160AB, GJ 3412AB, GJ 3916AB, 
Gl 644AB, Gl 735AB). For five of these stars, the components are well 
separated. For GJ 3412AB (observed with SOPHIE), the two components are 
blended. In this case we applied multi-Gaussian fits to the cross-correlation 
profile. 

\section{Sources of uncertainty in our $v\sin i$ values}

\subsection{Uncertainties due to the background}

% Fig. 3
\begin{figure*} 
\vspace{-1.5cm}
\begin{centering}
\includegraphics[width=14cm,angle=-90]{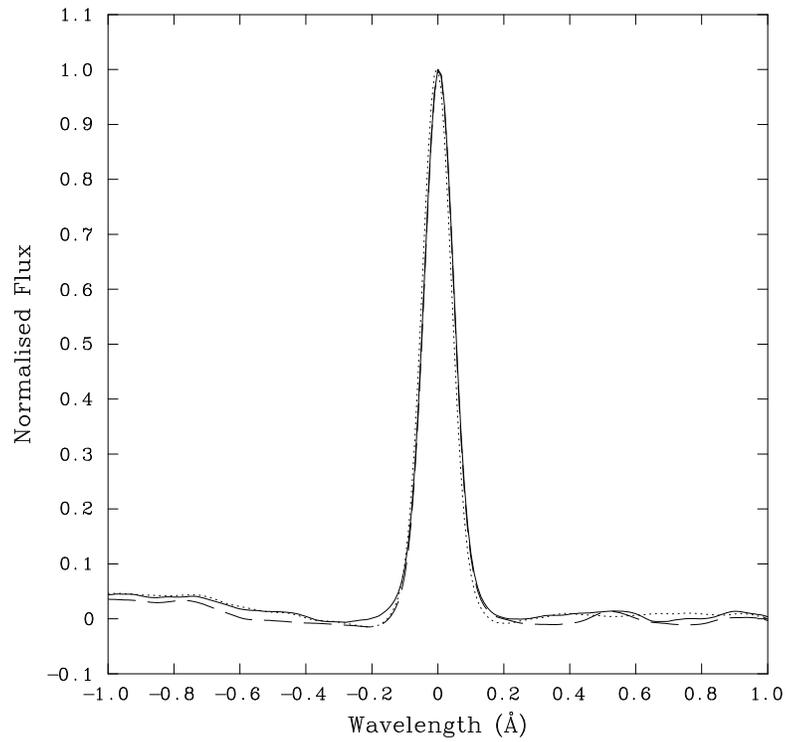}
\end{centering}
\vspace{-0.5cm}
\caption[]{Cross-correlation functions derived for GJ 4129 ($v\sin 
i = 0.50 km\ s^{-1}$, continuous line), Gl 581 ($v\sin i = 0.50 km\ s^{-1}$, 
dashed line) and Gl 781.1A ($v\sin i = 0.25 km\ s^{-1}$, dotted line) with 
GJ 1097, corrected for the background in GJ 1097.  The major source of 
uncertainty in our measurements of $v\sin i$ is caused by the (small) 
fluctuations in the background from one star to another.}
\end{figure*}

In Houdebine (2010b), using four different spectrographs and comparing our 
results with those of other authors, we showed that we could attain a 
detection limit of 1 $km\ s^{-1}$ and a precision of 0.3 $km\ s^{-1}$ for 
dM2 stars. Here, for dM3 stars the situation is very similar, except that we 
have even lower background noise in the cross-correlation functions than for 
dM2 stars. To illustrate this, we show the cross-correlation functions 
for three stars in Fig.~3: GJ 4129 ($v\sin i = 0.50 km\ s^{-1}$), Gl 581 
($v\sin i = 0.50 km\ s^{-1}$, a subdwarf), and Gl 781.1A ($v\sin i = 0.25 
km\ s^{-1}$). One can see in this diagram that the cross-correlation peaks 
for the three stars are very similar to one another, because of their low 
$v\sin i$. This is an indication of the excellent stability of the 
spectrographs we use. On the other hand, one can also note that there are 
small variations from star to star in the cross-correlation background. We 
estimate that the amplitude of these variations are about 2\% ($3\sigma$) of 
the intensity of the cross-correlation peaks. We found that this uncertainty 
on the background yields an uncertainty of $3\sigma$=0.0034\AA\ on the FWHM 
for HARPS, and $3\sigma$=0.0042\AA\ on the FWHM for SOPHIE. Therefore our 
measurements for dM3 stars are even more precise than those for dM2 stars. 
Moreover, these figures are estimates due to the noise in the background. But 
the S/N ratio in the cross-correlation peaks themselves is higher (several 
hundred). If our measurements were incorrect we would not obtain such good 
correlations: instead we would obtain  scatter diagrams. Further confirmation 
that our measurements are not incorrect is provided by the continuity in the 
gradients of the RAC's which are found to increase continuously from our dK5 
sample to our dM4 sample (Houdebine et al. 2015). 

As one can see from Fig.~2, the uncertainty of $3\sigma$=0.0034\AA\ on the 
FWHM yields a variable uncertainty on $v\sin i$ depending on its magnitude. 
Indeed, because the curves in Fig.~2 are increasingly steep at small 
$v\sin i$, the uncertainty in $v\sin i$ increases at low values of $v\sin i$. 
We quantify these uncertainties ($\pm\sigma$) on $v\sin i$ for each 
measurement for HARPS and SOPHIE and list the results in Table~2. If we take  
into account the limits of $\pm 3\sigma$ given above, the value of $v\sin i$ 
cannot be reliably measured below 0.95 $km\ s^{-1}$ for HARPS. The 
corresponding figure is 1.22 $km\ s^{-1}$ for SOPHIE. The width of the 
template cross-correlation function is only 4.61 $km\ s^{-1}$ for HARPS. A 
1 $km\ s^{-1}$ broadening corresponds to a broadening of the cross-correlation 
function of 3\%. We show the results of our convolved cross-correlations in 
Fig.~4 for values of $v\sin i$ of 0.0 (GJ 1097), 0.5, 1.0, 1.5, 2.0, 2.5, 3.0, 
4.0, 5.0, 7.0, 10.0 and 15.0 $km\ s^{-1}$ for HARPS. We obtain our simulations 
of rotational broadening of our cross-correlation profiles by convolving them 
with our rotational profiles. We find that we can measure $v \sin i$ values 
down to 250 $m\ s^{-1}$. Inspection of Fig.~4 shows that one can easily 
differentiate the template profile (left-most curve) from the profile which 
has been broadened by 1.0 $km\ s^{-1}$ (shown as the third curve in from the 
left-most curve). This confirms our previous findings that our 
cross-correlation technique enables us to evaluate $v\sin i$ readily down to 
values as small as 1.0 $km\ s^{-1}$. Further observations in the future with 
higher spectral resolution spectrographs such as ESPRESSO (Echelle 
SPectrograph for Rocky Exoplanet and Stable Spectroscopic Observations, ESO, 
R=220,000) could confirm our measurements below our $3\sigma$ detection limit.

% Fig. 4
\begin{figure*} 
\vspace{-1.5cm}
\hspace{-2.5cm}
\includegraphics[width=14cm,angle=-90]{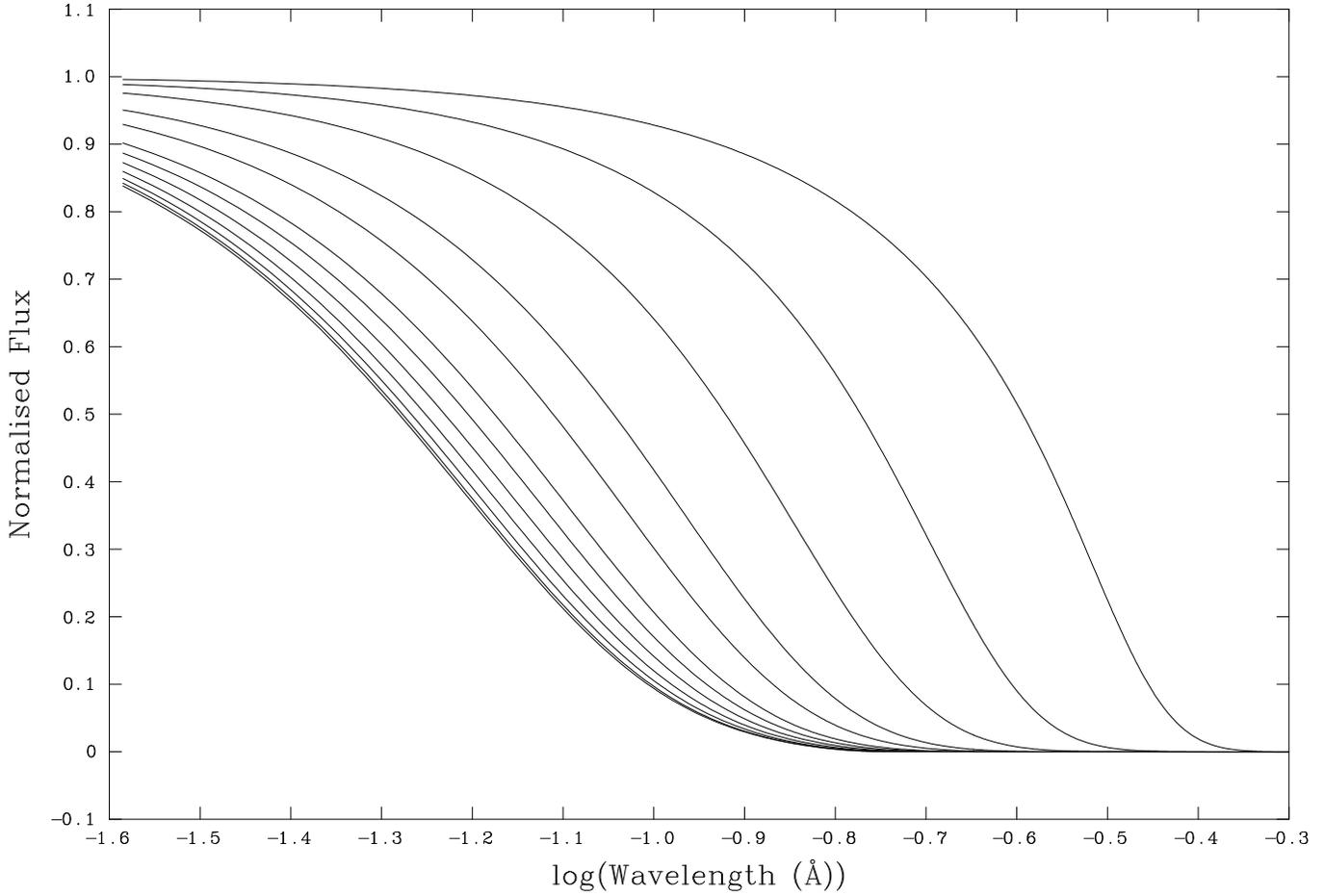}
\vspace{-0.5cm}
\caption[]{Cross-correlation function of GJ 1097 convolved with the rotational
profiles for $v\sin i$ values of 0.0 (left-most curve), 0.5, 1.0, 1.5, 2.0, 
2.5, 3.0, 4.0, 5.0, 7.0, 10.0 and 15.0 (right-most curve) $km\ s^{-1}$. By 
superposing actual stellar data on these curves, we find that we can measure 
the broadening of our template cross-correlation profile (GJ 1097 for HARPS) 
down to $v \sin i$ values of 0.25 $km\ s^{-1}$. Here the curve corresponding 
to our detection threshold of 1 $km\ s^{-1}$ is clearly separated in the 
diagram from the zero-rotation limit.}
\end{figure*}

\subsection{Gaussian profiles versus rotational profiles}

We also convolved our template cross-correlation profile for HARPS (GJ 1097) 
with Gaussians in which the broadenings were set equal to those of our 
theoretical rotational profiles (as listed above). Once again, we found that 
we could distinguish rotational broadening for $v\sin i$ as low as 250 
$m\ s^{-1}$. However, we found that in general the broadened profiles 
using Gaussians are narrower than those broadened with our rotational profiles 
taking into account limb-darkening effects. To illustrate this, we show in 
Fig.~5 the FWHM of our Gaussian broadened profiles as a function of the FWHM 
of our profiles broadened when taking into account limb-darkening. One can see 
that the Gaussian broadened profiles are narrower than the profiles that 
account for limb-darkening. There are almost no differences for low values of 
$v\sin i$ and large values of $v\sin i$. But for intermediate values 
$3<v\sin i<7 km\ s^{-1}$, the differences can be large: this difference can be 
as large as 0.9 $km\ s^{-1}$ for $v\sin i$ at about 4 $km\ s^{-1}$. Therefore 
it is important to take into account limb-darkening effects when measuring 
$v\sin i$ and $P/\sin i$.

% Fig. 5
\begin{figure*} 
\vspace{-1.5cm}
\hspace{-4.5cm}
\includegraphics[width=14cm,angle=-90]{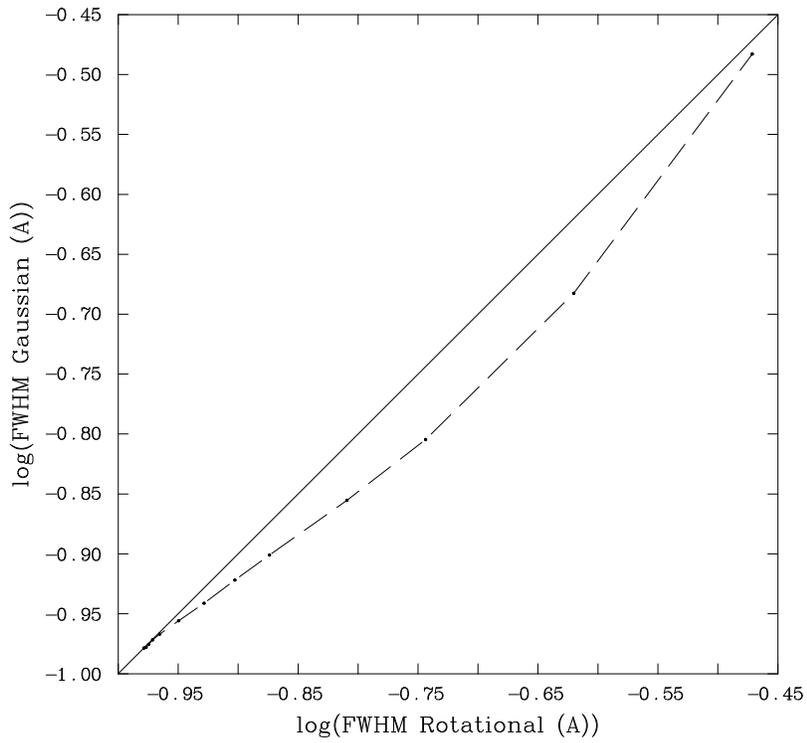}
\vspace{-0.5cm}
\caption[]{Dashed line: comparison of the FWHM of broadened profiles of 
GJ 1097 calculated by two different methods: (i) using our theoretical 
rotational profiles (see values on x-axis) and (ii) using Gaussian profiles 
(see values on y-axis). Solid line: results would lie along this line if both 
methods produced identical broadening. Along the dashed line, the Gaussian 
broadened profiles are systematically narrower than the theoretical rotational 
profiles.}
\end{figure*}

\subsection{Comparison with previous measurements}

In order to further assess the precision of the rotational velocity 
measurements reported in this paper, we searched the literature for other 
$v\sin i$ or rotational period $P$ measurements for our M0, M2, M3 and M4 
targets. We found several measurements of $v\sin i$ (Vogt et al. 1983, Marcy 
\& Chen 1992, Delfosse et al. 1998, Mohanty \& Basri 2003) and $P$ (Pizzolato 
et al. 2003, Kiraga \& Stepie\'n 2007, Irwin et al. 2011). We computed the 
$v\sin i$ from the rotation periods $P$ assuming $\sin i=1$. We plot in Fig.~6 
our best measurements versus the measurements from other spectrographs (e.g. 
SOPHIE as a function of HARPS, 7 measurements in the present sample of dM3 
stars, see Table~2, and 19 in total) and the measurements from other authors 
(46 measurements). One can see in this figure that the overall agreement is 
good, even for sub-$km\ s^{-1}$ measures (13 measures). (The solid line in 
Fig. 6 indicates where the results would lie if the agreement were perfect.) 
The mean of the differences between our best measures and other measures are 
0.42~$km\ s^{-1}$ and 1.13~$km\ s^{-1}$ for the slow rotators and the fast 
rotators respectively. This value for the slow rotators is presently lower 
than the $3\sigma$ estimated above due to the uncertainty in the background of 
the cross-correlations. Furthermore, this figure allows us to expect that we 
should be able to measure $v\sin i$ down to 0.50~$km\ s^{-1}$ for our M dwarf 
samples, i.e. a factor of 2 better than our estimates given above. The value 
of 1.13~$km\ s^{-1}$ for the fast rotators is much higher than our present 
estimates for these objects (see Table~2). Indeed, we should obtain the best 
precision for the fast rotators. The scatter of individual values in Fig. 6 
relative to the perfect agreement line can be partly explained as follows: i) 
limb-darkening effects were not taken into account by all authors in the 
derivation of $v\sin i$ values in the literature, and ii) we assumed 
$\sin i=1$ for some of the stars when in reality, the value of $\sin i$ must 
in some cases be smaller than 1. In fact, in the case of fast rotators, 
different authors report differences of about 1~$km\ s^{-1}$ between their 
observations of $v \sin i$: this might be due to the fact that these highly 
active stars are spotted, and that the presence of large spots which can 
appear and disappear between different observing epochs yields varying values 
of $v\sin i$. 

We note two large discrepancies between our measures and those of Kiraga \& 
Stepie\'n (2007) (not visible in Fig.~6). First, we found Gl 431 to be a rapid 
rotator $v\sin i=22.1 km\ s^{-1}$, in agreement with its strong CaII and 
H$_{\alpha}$ emissions (Papers XVII and XVIII) whereas Kiraga \& Stepie\'n 
(2007) give a rotation period of 14.31 days ($v\sin i\sim 1.5 km\ s^{-1}$). 
Obviously their detection of the rotational modulation seems spurious. ii) for 
Gl 669A we have $v\sin i=1.86 km\ s^{-1}$ whereas Kiraga \& Stepie\'n (2007) 
give a rotation period of 0.950 days ($v\sin i\sim 28.5 km\ s^{-1}$). 
Obviously, here again they have a spurious detection.

% Fig. 6
\begin{figure*} 
\vspace{-1.5cm}
\hspace{-4.5cm}
\includegraphics[width=14cm,angle=-90]{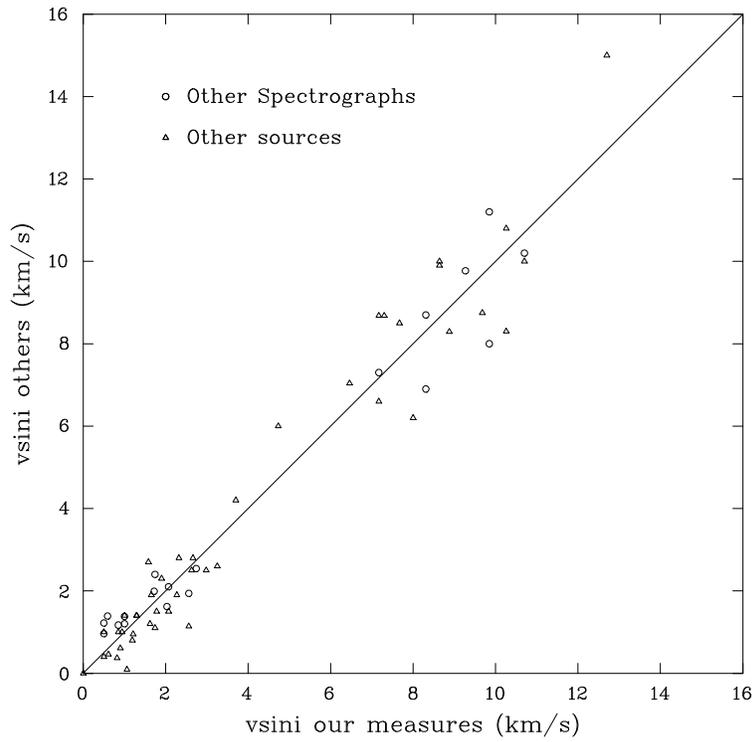}
\vspace{-0.5cm}
\caption[]{Comparison between our measurements of $v\sin i$ (x-axis) and 
measurements of $v\sin i$ from other spectrographs and from other authors 
(y-axis). Included in the plot are data for all of our sample M dwarfs 
(including subtypes M0, M2, M3 and M4). The solid line indicates perfect 
agreement between measurements.} 
\end{figure*}

\subsection{The distributions of the rotation periods}

We have compiled all of our rotation periods from the combined sub-samples of 
dM0, dM2, dM3 and dM4 stars: the total number of stars is 277. We sum them all 
and plot a histogram of the number of stars as a function of the rotation 
period in Fig.~7. Our histogram shows 3 peaks: (i) fast rotators at periods 
close to 5 days; (ii) the largest peak at periods close to 15 days; (iii) a 
weaker peak at periods close to 23 days. In order to estimate the validity of 
this histogram, we overplot the histogram of the rotation periods reported by 
Nielsen et al. (2013) for a sample of more than 12,000 M dwarfs. Nielsen et al.
 obtained their results by analyzing photometric data from the Kepler 
spacecraft: in the presence of long-lived starspots, photometry allows a 
direct measurement of the rotation period, without the uncertainties 
associated with $\sin i$. In view of the $\sin i$ factor, which cannot have a 
value in excess of unity, we expect that our histogram of $P / \sin i$ will be 
skewed towards longer periods in comparison with the measurements of Nielsen 
et al.  

Nevertheless, inspection of Fig. 7 indicates that the histogram of Nielsen et 
al. (2013) overlaps with ours in certain respects. Most importantly, the main 
peak occurs at about 15 days: in fact the median value of the Nielsen et al. 
histogram occurs at 15.4 days. This overlaps well with our largest peak. 
Moreover, a secondary (smaller) peak occurs in the Nielsen et al results at 
periods of 3-5 days, overlapping with our "fast rotator" peak. At longer 
periods, i.e. among the slow rotators, our distribution has a longer tail than 
the Nielsen et al. histogram, but this behaviour is to be expected based on 
the $\sin i$ factor in our data.  

A separate investigation of rotational periods among a large sample of M 
dwarfs has also been reported by McQuillan et al. (2013), also using Kepler 
data. In a sample of 1570 M dwarfs, the distribution of rotation periods is 
found to be bimodal, with peaks at about 19 and 33 days. McQuillan et al use 
proper motion data to suggest that the two peaks correspond to stars of 
different ages, with the older group having the longer rotation periods. This 
interpretation suggests that the slower rotating stars may belong to the halo 
population, whereas the faster rotations may belong to the disk population. As 
we noted above (Section 2.2), in the present study, we calculate the period 
values (and the period distribution) ${\it for disk stars only}$: we do not 
include halo stars. This suggests that our sample of periods should include 
few (if any) halo stars: in view of this, we would not expect to recover the 
peak at 33 days reported by McQuillan et al. Only the 19 day peak in the 
McQuillan et al results should be compared to our results: this presumably 
corresponds to the 15 day peak in our data. The difference between the peak 
periods (19 vs 15 days) may be associated with the much larger sample of stars 
observed by McQuillan et al: the latter sample is larger than ours by a factor 
of 5-6, and may contain more halo stars than our sample.    

Overall, allowing for significant differences in sample sizes, our results 
seem to agree rather well with the more precise results obtained from the 
rotational modulation for the larger number of M dwarfs which have been 
observed with Kepler. 

% Fig. 7
\begin{figure*} 
\vspace{-1.5cm}
\hspace{-4.5cm}
\includegraphics[width=14cm,angle=-90]{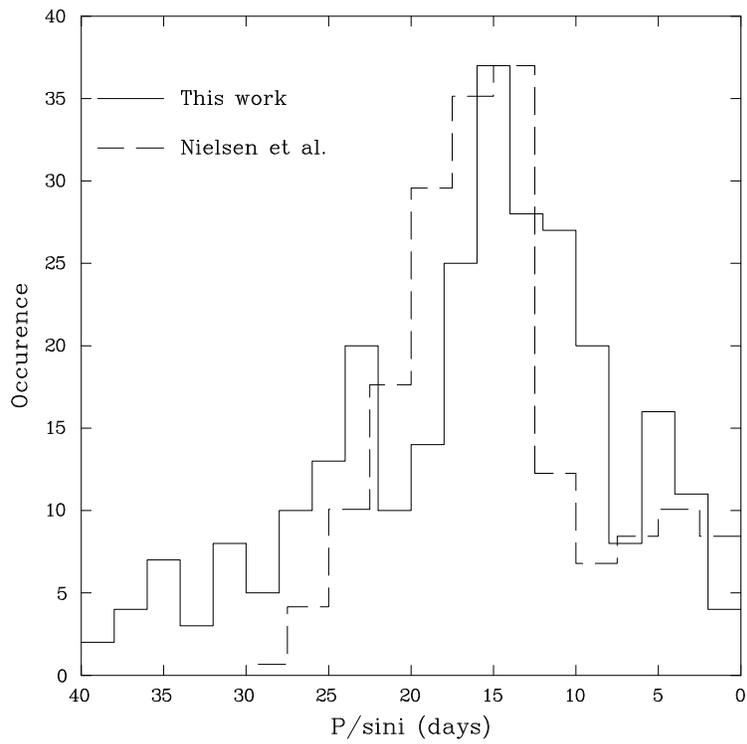}
\vspace{-0.5cm}
\caption[]{Solid curve: Number of stars in our sample of M dwarfs as a 
function of the rotation period. Included are all M dwarfs from the combined 
data for sub-types M0, M2, M3 and M4. Dashed curve: results of Nielsen et al. 
(2013) for a much larger sample of stars. }
\end{figure*}

The reason why we can attain precisions as good as those mentioned above in 
our dM3 stars is the same as for our earlier work on dM2 stars: our 
cross-correlation profiles have a very high S/N ratio (several hundred: see 
Fig. 3). This is due to the fact that the cross-correlations are based on 
combining the profiles of many spectral lines. It is therefore in theory 
possible to measure subtle differences between different cross-correlation 
profiles (see Fig.~4) of only a few percent. Our results can be compared to 
the noisier background in Reiners et al. (2012) where background noise is 
about 20% of the main cross-correlation peak. It is this large noise level 
which limits their detection ability to $v\sin i$ values in excess of 
3~$km\ s^{-1}$.

\section{Rotation of M3 dwarfs}

% Fig. 8
\begin{figure*} 
\vspace{-0.5cm}
\hspace{-4.5cm}
\includegraphics[width=14cm,angle=-90]{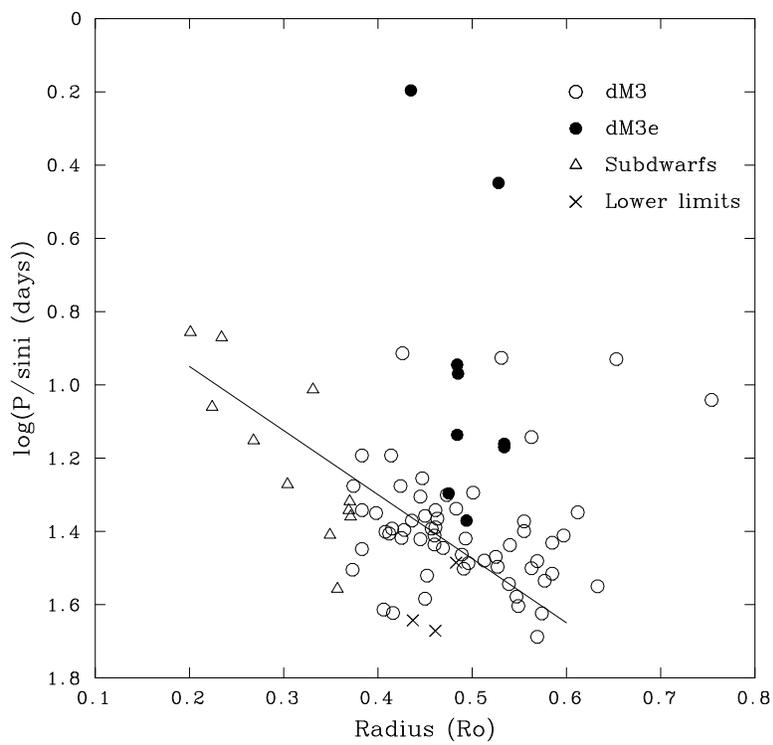}
\vspace{-0.5cm}
\caption[]{
$P$/sin$i$ measurements as a function of stellar radius. Open circles: dM3 
stars; filled circles: dM3e stars; triangles: sdM3; crosses: lower limits.}
\end{figure*}

In Fig.~8, the projected rotation periods $P$/sin$i$ are shown as a function 
of stellar radius. (Note the inverted scale on the vertical axis.) In order to 
to highlight differences between the less active (dM3) stars, the more active 
(dM3e) stars, and the subdwarfs (sdM3), we use different symbols for these 
different types of M3 dwarfs in Fig. 8. The solid line through dM3 and sdM3 
data  illustrates that $P$/sin$i$ tends to decrease (from 45 days to about 9 
days) as the radius decreases (from 0.6 R$_{\odot}$ to 0.2 R$_{\odot}$. Such a 
trend was also found in our sample of dM4 stars, and also to some extent in 
our dM2 stellar sample. In our sample of dM3 stars, we have only a few 
subdwarfs, but the few which are in the sample do indeed have on average 
shorter rotation periods than the main sequence stars do. Our data 
indicate that, in spite of their probable older age, subwarfs rotate more 
rapidly than normal dwarfs. Although the results of Savcheva, West \& 
Bochanski (2014) suggest that M type subdwarfs are not members of the halo 
unless they are "extreme" or "ultra" subdwarfs. Our results highlights the 
fact that magnetic braking mechanisms among early M dwarfs depend mainly on 
the stellar radius and mass (Houdebine et al. 2015); the magnetic braking 
mechanisms decrease more rapidly in efficiency with stellar radius than their 
efficiency integrated over age.

The study of the distributions of stellar rotation periods as a function of 
spectral type is particularly important for our understanding of stellar 
rotation dynamics. These distributions may provide essential new constraints 
for the mechanisms of magnetic braking. Fig.~9 illustrates the histogram of 
$P$/sin$i$ values which we have obtained in this paper for our sample of dM3 
stars. In separate panels, we also present, for purposes of comparison, the 
corresponding histograms for our samples of dM2 and dM4 stars (data for which 
were reported in earlier papers). Inspection of the dM3 histogram indicates 
that our sample includes projected periods as long as 40-45 days, and as short 
as 1-2 days. It appears that the dM3 histogram contains one well defined peak, 
and perhaps also some smaller peaks. However, the statistics of small numbers 
may lead to only marginal significance for the smaller peaks. The dM3 
histogram also contains several fast rotators: the two fastest rotators in our 
sample, Gl 896A and Gl 207.1 (with $v\sin i$ = 14 and 9.5 $km ~s^{-1}$) are 
both dM3e stars. Thus, the stars which exhibit the fastest rotation are also 
stars with a high level of magnetic activity. This is consistent with the RAC. 
We consider the dM3 stars in our sample as slow rotators, while the dM3e stars 
are fast rotators, i.e. we consider the boundary where H$_{\alpha}$ starts to 
go into emission as equivalent to a boundary between slow and fast rotation 
among M3 dwarfs. 

Comparison of the dM3 and dM2 histograms indicates that our dM3 distribution 
extends over a broader range of periods than our dM2 distribution, despite the 
fact that their spectral classifications differ by only one decimal sub-type. 
For dM2 stars we observe that the distribution for slow rotators is rather 
compact, with the main peak around 11 days, and almost no stars at periods 
longer than 18 days. In contrast to this, in the dM3 histogram, the largest 
peak is attained at a period of 23 days and there are numerous stars with even 
longer periods. 
 
Comparison between the dM3 histogram and the dM4 histogram reveals even more 
striking differences between stars which also differ by only one decimal 
sub-type. For dM4 stars, the distribution is extremely compact with a maximum 
around 14 days, with possibly a second maximum around 6 days, and no stars at 
all with periods longer than 18 days. 

In view of these differences among the histograms, the results in Fig. 9 
suggest that rotational properties of M dwarfs do not vary in a manner which 
could be described simply as monotonic as we move from dM2 through dM3 to dM4 
spectral sub-types. Our results indicate that our sample of dM3 stars contains 
many stars which have rotation periods which are clearly longer than those of 
any of the dM2 and dM4 stars in our samples.

To quantify the differences between rotational properties of the various 
decimal sub-types of dM stars, we consider the average (projected) rotation 
period $P/\sin i~(ave)$ of the stars in our various sub-samples of slow 
rotators (i.e. excluding spectroscopic binaries, subdwarfs, and active 
chromosphere stars). For dM3 stars, including all of the inactive stars in our 
sample (numbering 49 stars), we find that averaging of the individual 
$P/\sin i$ measurements leads to $P/\sin i~(ave)$ = 25.8 days. If we include 
only dM3 stars out to a maximum distance of 13 pc (where our sample is closer 
to complete), we find $P/\sin i~(ave)$ = 23.5 for a sample of 25 stars (i.e. 
still excluding spectroscopic binaries, sdM3 and dM3e stars). These two values 
are very close, within $\pm 1.1$ days of an average of 24.65 days. Therefore 
the bias in our sample towards the brightest dM3 stars does not appear to have 
a significant selection effect on the value of $P/\sin i~(ave)$. In order to 
place an error bar on our estimates of $P/\sin i~(ave)$, we first calculate an 
individual error for the $P/\sin i$ value for each of our 49 dM3 stars. To do 
this, we include the errors on the absolute magnitudes and in the radius 
measurements (Table~1), and we also include the errors on the $v\sin i$ 
measurements (Table~2). Combining these errors, we derive the errors listed in 
Table 2 for each of the individual $P/\sin i$ measurements for our 49 stars. 
We then calculated the mean of these individual errors: we found the mean 
errors below and above the mean to be $\sigma = -5.98 days$ and 
$\sigma = +13.8 days$. Using these mean errors for the individual values of 
$P/\sin i$ for 49 stars, we obtained a mean error on the average 
$P/\sin i~(ave)$ for 49 stars by dividing the mean individual errors by 
$\sqrt{49}$. In this way, we find errors on $P/\sin i~(ave)$ ranging from 
-0.85 days to +1.97 days. Treating these as standard deviations on 
$P/\sin i~(ave)$, we conclude that the $3\sigma$ error bars on our computed 
values of $P/\sin i~(ave)$ are -2.56 days and +5.9 days. These error bars are 
included in Fig. 10 in association with the plotted value of $P/\sin i~(ave)$ 
= 25.8 days for dM3 stars. 

Turning now to our samples of dM2 and dM4 stars, we find that the mean 
projected period $P/\sin i~(ave)$ has values of 14.4 days and 11.4 days 
respectively. In order to assign error bars to these values, we considered the 
average $v\sin i$ and average radius for our subsamples of dM2 and dM4 stars. 
An upper value of the error on the mean periods is then given by the standard 
error on $v\sin i$. We found that the mean $v\sin i$ is 2.41 $km\ s^{-1}$ and 
1.57 $km\ s^{-1}$ for dM2 and dM4 stars respectively in our samples. The mean 
radii are 0.623 R$_{\odot}$, 0.490 R$_{\odot}$ and 0.366 R$_{\odot}$ for dM2, 
dM3 and dM4  stars respectively. These yield uncertainties of $\pm$1.44 days 
and $\pm$1.89 days on the $P/\sin i~(ave)$ values for dM2 and dM4 stars 
respectively. Along the same lines, we found that the mean radii are 0.815 
R$_{\odot}$ and 0.711 R$_{\odot}$, and the mean $v\sin i$ are 1.62 $km\ s^{-1}$
 and 2.28 $km\ s^{-1}$ for dK5 and dM0 stars respectively. This gives 
uncertainties of $\pm$6.00 days and $\pm$2.84 days for $P/\sin i~(ave)$ in dK5 
stars and dM0 stars respectively. The values of $P/\sin i~(ave)$ and their 
error bars are plotted for our samples of dK5, dM0, dM2, dM3, and dM4 stars in 
Figure 10.

On an observational selection note, we point out that some of our detections 
of $v\sin i$ lie below the $3\sigma$ threshold of 0.95 $km\ s^{-1}$ for HARPS 
and 1.22 $km\ s^{-1}$ for SOPHIE. As a result, the values of $P/\sin i~(ave)$ 
we give above are lower limits to the real figure of the average rotation 
period for dM stars. However, we see no reason why this should affect dM3 
stars more than dM2 or dM4 stars. Therefore, there should be no significant 
effect on the principal conclusion which we reach in this ${\it differential}$ 
study between neighbouring spectral sub-types: the value of $P/\sin i~(ave)$ 
for dM3 stars is abnormally large compared to the values for dM2 and dM4 stars.

If there existed a monotonic decrease in $P$/sin$i ~(ave)$ as the spectral 
type moved to later values, then given the values of 14.4 days and 11.4 days 
for dM2 and dM4 stars, one would expect dM3 stars to have an average rotation 
period of 12.9 days. Instead of that, our results indicate that dM3 stars 
rotate more slowly than this expectation, by an amount 25.8-12.9 = +12.9 days. 
Compared to our estimate above of the standard deviation of +1.97 days 
(towards longer periods), the deviation of dM3 stars from monotonic behaviour 
between dM2, dM3 and dM4 has a statistical significance of (6-7)$\sigma$. This 
suggests that there is something happening at spectral type dM3 which deviates 
significantly from the monotonic behaviour demonstrated by K5, M0, M2, and M4 
dwarfs. 

We conclude that on average, dM3 stars rotate slower than expected when 
compared with stars of immediately adjacent spectral sub-type. Moreover, the 
period distribution of dM3 stars exhibits a stronger skewing towards longer 
periods, with some members of the sub-sample rotating as slowly as 40-45 days. 
These slow rotations are consistent with the low activity levels of dM3 stars 
as measured in the Ca\,{\sc ii} lines (Houdebine et al. 2015). 

So far, we have discussed the rotational properties of the slow rotators among 
our dM2, dM3, and dM4 samples. Now let us turn to the more active stars, i.e. 
dM2e, dM3e, and dM4e, and ask: is there any evidence to indicate that the 
unusually slow rotation of M3 stars is also present among the active stars?
To address this topic, we have analysed independently the mean rotation 
periods for active stars (i.e. M dwarfs with H$_{\alpha}$ in emission). We 
find $P$/sin$i ~(ave)$ = 4.22$\pm$0.26, 2.82$\pm0.19$, 4.11$\pm$0.20, 
10.3$\pm$1.34 and 6.32$\pm$0.30 days for dK5, dM0, dM2, dM3 and dM4 stars 
respectively. The value for dM0 stars is not trustworthy, because of the 
smallness of the sample. The numerical values of $P$/sin$i~(ave)$ for other 
sub-samples (dK5e, dM2e, dM3e, dM4e) suggest that for the fast rotators, 
$P$/sin$i ~(ave)$ tends to increase towards later spectral types. This is the 
opposite trend to that of the slow rotators. One can explain this if one 
considers that dM4e stars stay longer at a high level of activity than dK5e 
stars (see RM). In other words, on average, the sample of active stars gets 
older as the spectral type increases (RM). Despite this behaviour, our 
averages indicate that dM3e stars once again stand apart from immediately 
adjacent decimal subtypes in the following sense: the dM3e stars have a 
significantly longer rotation period. 

Thus, the relatively long rotation period of M3 stars (relative to M2 and M4) 
emerges in our data for both slow rotators (dM3) and fast rotators (dM3e). 
Thus, whether we consider dM3e stars (relatively young) or dM3 stars 
(relatively older), some process of magnetic braking has slowed M3 stars down 
more efficiently than the adjoining M2 and M4 stars.

% Fig. 9
\begin{figure*} 
\vspace{-0.5cm}
\hspace{-4.5cm}
\includegraphics[width=8cm,angle=-90]{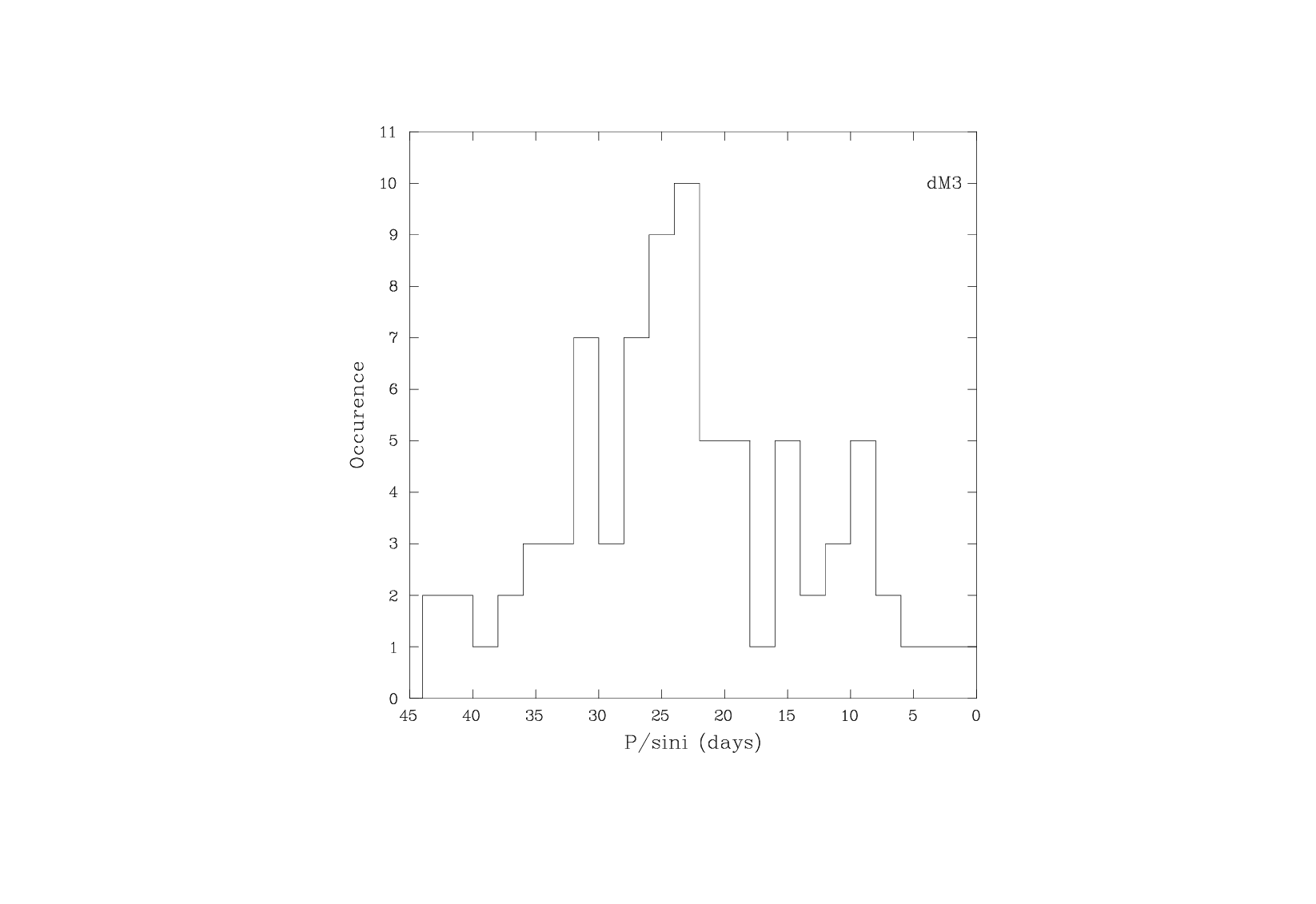}
\hspace{-4.5cm}
\includegraphics[width=8cm,angle=-90]{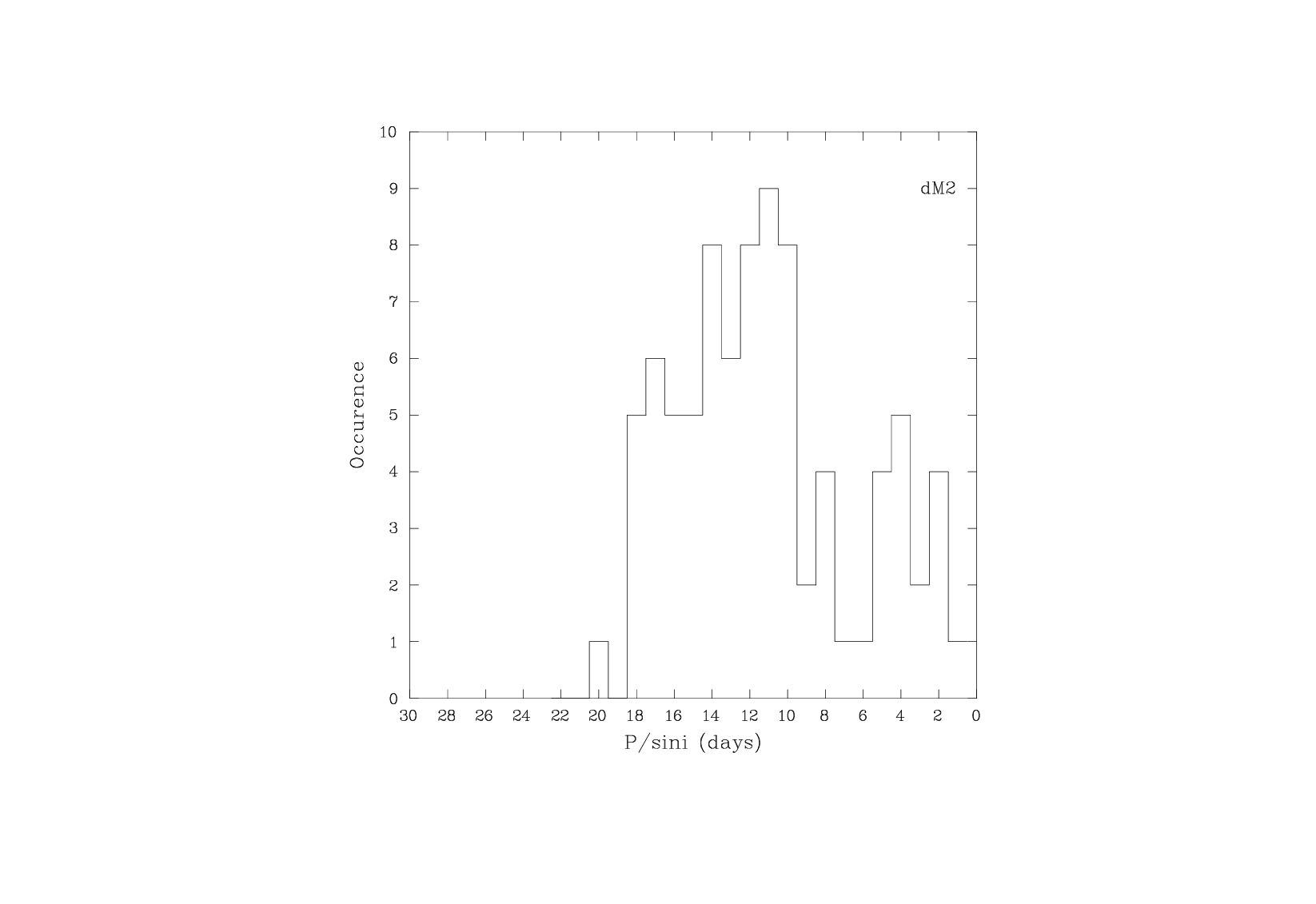}
\includegraphics[width=8cm,angle=-90]{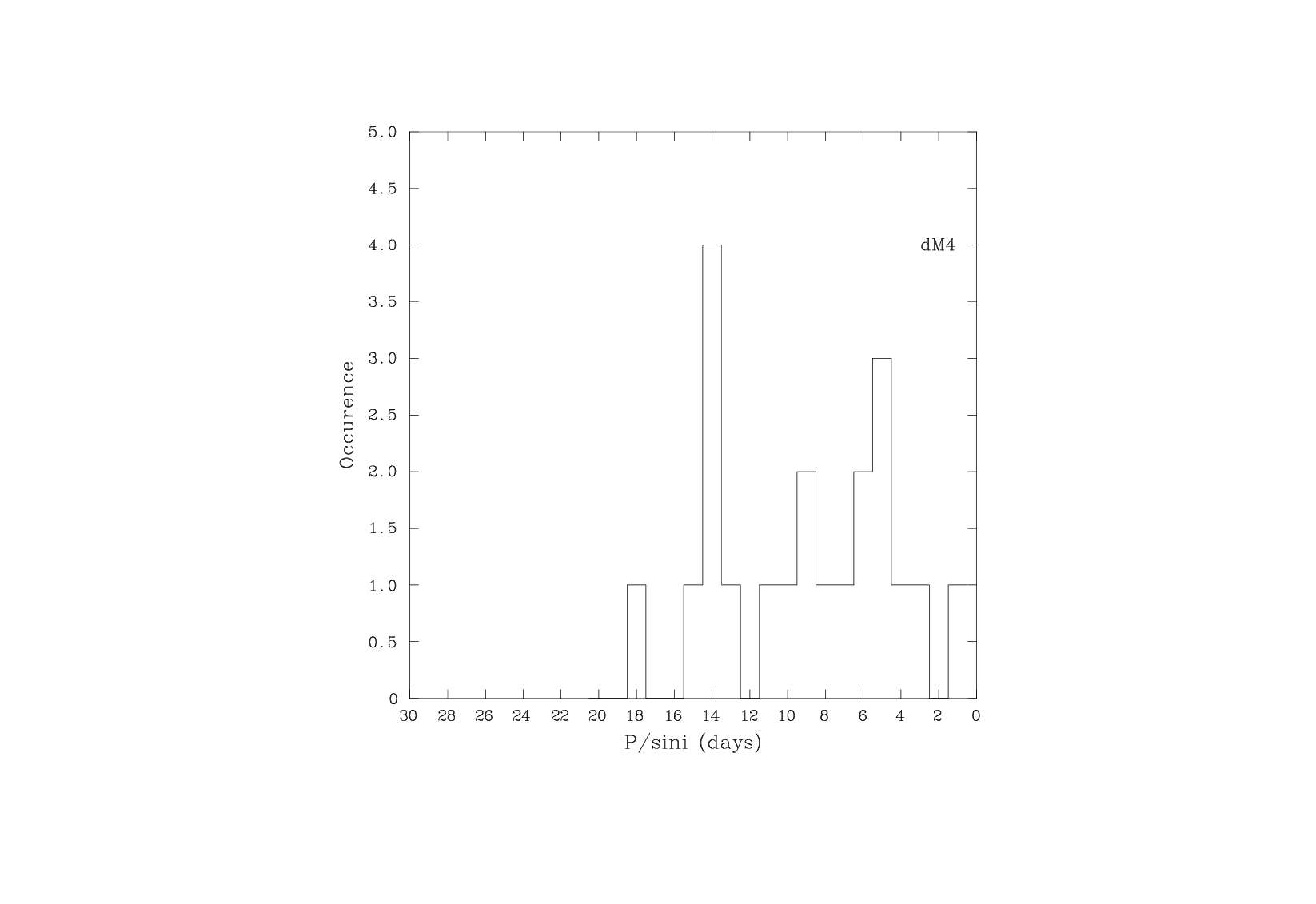}
\vspace{-0.5cm}
\caption[]{Histograms of $P$/sin$i$ for stars of spectral sub-types dM3, 
dM2, and dM4. Note that dM3 stars have rotational periods which extend to 
larger values than dM2 stars and dM4 stars. }
\end{figure*}

A summary of our results is illustrated in Fig.~10, where the mean rotation 
period $P/\sin i~(ave)$ is plotted as a function of (R-I)$_{c}$, with labels 
attached to the appropriate spectral sub-type. Fig. 10 indicates that, in an 
overall sense, there is a trend in $P$/sin$i ~(ave)$ to decrease from longer 
than 30 days to about 10 days as the spectral sub-type increases from dK5 to 
dM4. But a departure from this trend occurs among the dM3 stars: there, the 
value of $P$/sin$i ~(ave)$ rises to a peak which stands well above the value 
one might expect from a simple interpolation between dM2 and dM4. The slowest 
rotators in Fig. 10 are the dK5 stars with a mean rotation period longer than 
30 days. (This value is actually a lower limit because the $v\sin i$ values of 
many dK5 stars were below the detection limit of 1 $km ~s^{-1}$.) Overall, the 
long periods of dK5 stars indicate that rotational braking has been quite 
efficient for dK5 stars, whereas the short periods of dM4 stars indicate that 
rotational braking has {\it not} been as efficient for those stars. The dM3 
stars do not fit into this trend: the dM3 data suggest that at dM3, the 
braking mechanism is {\it more effective than the data for dM2 and dM4 stars 
would lead us to expect}.
 
The results in Fig~10 may provide an important constraint on the rotational 
braking mechanism, in particular at the fully convective threshold. In the 
next section, we turn to a discussion of this topic. 

% Fig. 10
\begin{figure*} 
\vspace{-0.5cm}
\hspace{-4.5cm}
\includegraphics[width=14cm,angle=-90]{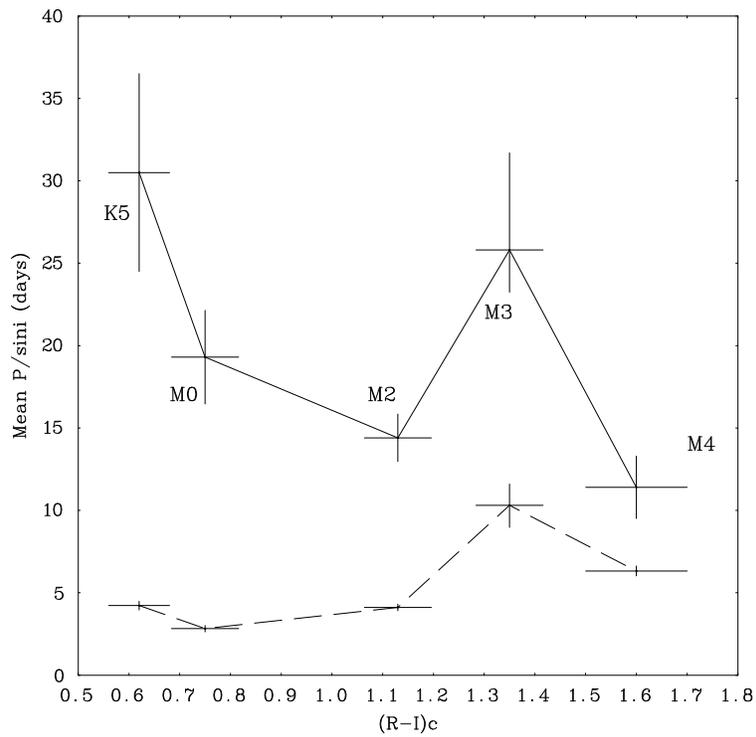}
\vspace{-0.5cm}
\caption[]{Mean $P$/sin$i$ values of inactive (continuous line) and active 
(dashed line) dwarfs as a function of (R-I)c. Note that, although there is an 
overall trend to shorter periods as spectral type increases, there is a 
significant departure from this trend at dM3. Theoretically, M3 is where 
transition to complete convection is thought to occur.}
\end{figure*}

\section{Discussion}

The results in Fig. 9 and Fig. 10 suggest that something unusual happens to 
the rotational properties of dwarfs at dM3, presumably associated with a 
change in $dJ/dt$. In this section, we review the quantities which determine 
$dJ/dt$ in order to ask: what might be unusual about M3? We shall discuss 
several of the factors mentioned by RM as worthy of attention (see Section 
1.4). 

\subsection{Contributions to dJ/dt: $\dot{M}$}

Two factors determine $dJ/dt$ (see Section 1.2.2): $\dot{M}$ and the “lever 
arm” length $D$ = $R_A$ (the Alfvenic radius) of enforced co-rotation. As 
regards $\dot{M}$, there are no reliable direct measures for M dwarfs. 
However,indirect information about $\dot{M}$ comes from studies of a "hydrogen 
wall" which accumulates in the interaction between the wind and the 
interstellar medium. Wood (2006) has analyzed the available data and concludes 
that in the M dwarfs, $\dot{M}$ may be smaller than the solar value. 

\subsection{Contributions to dJ/dt: the "lever arm" $D$}  

In all stars, $D$ can certainly not be $< R_*$. But $D$ may be $\gg R_*$ if 
the $B$ field is extended and strong. The larger $D$, the larger $dJ/dt$ for 
a given $\dot{M}$. In the solar wind, scintillation data suggest 
$D~\approx$ (15-20)$R_{\odot}$ (e.g. Mullan 1990). In M dwarfs, $D$ is very 
uncertain because $B$ is difficult to measure $\it in the wind$. On the $\it 
surface$ of an M dwarf, the large-scale fields $B$ are as strong as 1-4 kG 
(Saar, 1996; Reiners et al 2009), i.e. 100 times stronger than the large-scale 
(polar) fields in the Sun. As a result, we expect that $D$ in M dwarfs may 
greatly exceed the solar value, thereby contributing to larger $dJ/dt$. 

However, it is not merely the strength of $B$ at the surface which contributes 
to $D$: the radial behaviour $B(r)$ (i.e. the topology) is also important.
Some M dwarfs are found to have dipole (global) topology (Donati et al. 2006), 
whereas others have small-scale (multipole, active regions) topology (Morin et 
al. 2010). Multipole fields of degree $\it l$ fall off in strength with 
increasing radial distance as $1/r^{2+l}$. Dipoles ($l$ = 1) fall off as 
$1/r^3$, while quadrupole ($l$ = 2) and octupole ($l$ = 3) fall off $1/r^4$ 
and $1/r^5$. Thus, multipoles are so strongly radial dependent that they tend 
to "close in on themselves" quite close to the surface of the star. Thus, 
multipoles have smaller "lever arms" than dipoles. Other things being equal, 
$dJ/dt$ is smaller in a multipole field. To understand why some M dwarfs 
possess dipolar fields, while others possess multipolar fields, Gastine et al 
(2012) have used a dynamo model in a deep convective envelope to show that 
fast (slow) rotators tend to favor dipole (multipole) fields. In the present 
paper, with our deliberate choice of slow rotators, we should be dealing 
mainly with multipolar fields, i.e. stars where the sizes of active region 
loops will play a role in determining the lever arm $D$. 

When we discuss magnetic topology in M dwarfs, we need to ask: does the 
topology change from one side of the TTCC to the other? Shulyak et al (2014) 
have discussed this issue by observing 4 stars near the TTCC: but they reported
 "no difference between the field distribution of partially and fully 
convective stars". We will discuss this result below (Section 7.6).

\subsection{The value of $D$ and its relationship to magnetic loop lengths}

In the presence of multipole fields, it seems plausible that the value of $D$ 
may be determined by the length of the largest closed loops. In the solar 
corona, X-ray images indicate that the maximum height of closed loops is 
$\leq (0.1-0.5)R_{\odot}$. Can we estimate loop lengths in M dwarfs? Mullan et 
al. (2006) have reported a study of flaring loops in which the parameters of 
flare light curves allow extraction of loop lengths. In M dwarfs with colours 
which are redder than a certain limit, the loop lengths (expressed in terms of 
$R_*$) were found to be several times larger than in solar-like stars. The 
onset of larger loops in cooler M dwarfs, with the associated increase in $D$, 
would contribute to an {\em increase} in $dJ/dt$, i.e. an increase in the 
ability to slow down stellar rotation to longer periods. Might this be related 
to the abnormal rotational properties of dM3 stars which we  have identified 
in the present paper? The answer depends on precisely where the increase of 
loop lengths occurs in M dwarfs. 

\subsection{At what spectral type do large loop lengths set in among M dwarfs?}

Mullan et al (2006) report that long loops set in at B-V = 1.4-1.5. 
Unfortunately, B-V values change so slowly between M0 and M4 that the above 
B-V range may lie anywhere between M0 and M4. However, Mullan et al. (2006) 
also report that long loops set in at V-I = 2.1-2.3. According to the online 
website http://www.stsci.edu/~inr/intrins.html the values of V-I at M2, M3, 
and M4  are 2.06, 2.24, and 2.43. Thus the onset of long loops occurs later 
than M2 and earlier than M4. The best fit is M3. 

\subsection{TTCC and dynamo activity}

Dynamo activity in low mass stars is expected to change its properties at the 
TTCC. In stars with radiative cores, an interface dynamo can operate. But 
when the star is completely convective, such a dynamo is impossible. This has 
long suggested that the dynamo should undergo a change of regime across the 
TTCC, perhaps to a distributed dynamo (e.g. Mullan and MacDonald 2001). 
Surprisingly, no evidence for such a change can be seen in the best X-ray data 
(Mullan and MacDonald 2001). But perhaps the X-ray data is not the best place 
to look for a change of dynamo regime. Where else might we look?

\subsection{TTCC and loop lengths}

Since the increase in loop length reported by Mullan et al (2006) occurs at 
M3, and this is the spectral type associated with the theoretical prediction 
of the TTCC, we suggest that an increase in loop length is associated with a 
change in dynamo regime at the TTCC. Even if $\dot{M}$ did not change from M2 
to M3, an increase in loop length in going from M2 to M3 would result in a 
sudden increase in $D$, with a concomitant increase in $dJ/dt$. This would 
cause more effective braking of stellar rotation at M3 at the TTCC compared to 
earlier M dwarfs, with their smaller loops. This could explain why the mean 
$P/\sin~i$ in Fig. 10 is longer for M3 stars than for M2 stars.

What about M dwarfs later than M3? They also have large loop lengths (Mullan 
et al. 2006), so should they not also have long rotation periods?  The answer 
would be yes, if $\dot{M}$ were the same at M4 as at M3. However, the (scant) 
data available from the study of Wood (1996) hints that $\dot{M}$ may tend 
towards smaller values at later spectral types. To the extent that this is 
true, the magnitude of $dJ/dt$ would become smaller at spectral type M4 even 
if the loop lengths remained as large as those at M3. In the presence of 
reduced $dJ/dt$, the M4 stars would be expected to rotate faster than M3 
stars. This is consistent with the results which were shown above in Fig.~10.

Finally, if M3 is indeed the location of the TTCC, we can understand better 
the conclusions of Shulyak et al (2014), who reported that they found "no 
difference between the field distributions" in their sample of 4 stars. The 
fact is, their target stars were all of spectral type M3.5 or later, i.e. we 
would classify all 4 stars as fully convective. This could explain why all 4 
stars have similar field distributions.

\section{Conclusion}

In this paper, we have presented values of projected rotational velocity 
$v \sin~i$ for a sample of 82 stars belonging to the narrow spectral sub-type 
dM3. The velocities were extracted from archival spectra which had been 
obtained by observers using two spectrographs designed to be among the 
stablest and highest resolution instruments currently available. These 
instruments allowed us to determine values of $v \sin ~i$ down to, or even 
somewhat below, 1 $km~s^{-1}$. In order to achieve such high resolution and 
precision, we used a cross-correlation technique which fits the line profiles 
of hundreds of spectral lines. In earlier papers by one of us (E.R.H.), the 
cross-correlation technique has also been used to obtain $v \sin ~i$ for stars 
of spectral sub-types dK5, dM2, and dM4. Comparison of our $v \sin ~i$ values 
with results obtained by other observers for overlapping targets shows 
excellent agreement. 

Combining our values of $v \sin ~i$ with stellar radii for each star, we 
obtain a "projected rotational period" $P/\sin~i$ for each star in our dM3 
sample. A histogram of $P/\sin~i$ for our dM3 sample spans a range from as 
short as 1-2 days to as long as 40-45 days, with a peak at 23 days. For the 49 
slowly rotating dM3 stars in our sample (i.e. excluding dM3e stars, subdwarf 
sdM3 stars, and spectroscopic binaries), we find that the mean $P/\sin~i$ is 
$P/\sin~i ~(ave)$ = 25.8 days. We estimate that the $3 \sigma$ uncertainties 
in this estimate are (-2.56, +5.9) days.
 
In view of the availability of earlier papers on rotational properties of 
several spectral sub-types, we have undertaken in this paper a detailed 
comparison between the rotational properties of dM3 stars and those of dK5, 
dM0, dM2, and dM4. We find that the histograms of $P/\sin~i$ for our samples 
of dM2 and dM4 stars are considerably narrower than for our dM3 sample. In 
particular, the dM3 rotational distribution contains a larger number of slow 
rotators than the neighbouring sub-types. Quantitatively, we find that the 
values of $P/\sin~i ~(ave)$ for dM2 and dM4 stars are 14.4 and 11.4 days 
respectively. Linear interpolation between these two values suggests that dM3 
stars might be expected to have $P/\sin~i ~(ave)$ = 12.9 days: but this 
expectation is shorter by several $\sigma$ than the value we have actually 
derived for $P/\sin~i ~(ave)$ in dM3 stars. We conclude that the dM3 stars in 
our sample have been subjected to braking which is significantly more 
effective than the braking which has occurred in the adjacent spectral 
sub-types dM2 or dM4. We can think of no observational selection effect which 
would cause our dM3 sample to be spuriously affected in this way relative to 
our dM2 and dM4 samples: all samples were chosen from the same sets of 
archival data. 

In an attempt to understand why braking of stellar rotation might be different 
at dM3 than at dM2 or dM4, we note that magnetic loop lengths obtained from 
analysis of flare light curves (Mullan et al. 2006) increase to longer values 
at particular values of B-V and V-I colours. Interestingly, the V-I colour at 
which the loop sizes increase turns out to be spectral sub-type M3. To the 
extent that these two factors (rotation, loop length) are not merely 
coincidental, we regard the occurrence of both at sub-type M3 as an important 
conclusion of the present paper. 

Especially significant is the fact that stellar structure theory predicts that 
stars on the main sequence should make a transition to complete convection 
(TTCC) at spectral type M3. It has often been speculated that a change in 
dynamo regime probably occurs at the TTCC: this could explain why loop lengths 
change at M3. In this context, we have proposed a possible explanation 
(Section 7.6) why the rotational properties of dwarfs might exhibit an 
abnormality at M3 relative to adjacent spectral sub-types. 

The speculated switch in dynamo regime at the TTCC might be testable if 
further observational evidence were to reveal some other unusual property of 
dM3 stars relative to their neighbours (dM2 and dM4 stars). One possibility 
for such a test involves the frequencies of p-modes. It is known that magnetic 
fields inside a star cause structural changes which result in systematic 
shifts of p-mode frequencies (e.g. Mullan et al. 2007). A star with an 
interface dynamo (ID) located deep in the interior of the star might give rise 
to different magnetically-induced structural changes from those in a star with 
a distributed dynamo (DD) where fields are generated by small-scale turbulent 
processes throughout the convection zone (Durney et al. 1993). If the 
differences in structural changes between ID and DD turn out to be large 
enough, a study of frequency shifts among p-modes in dM2, dM3 and dM4 stars 
might help to strengthen the case for a switch in the dynamo regime at M3.  

\section*{acknowledgements}
The authors thank an anonymous referee for detailed and constructive comments 
which have helped to improve the paper. This research has made use of the 
SIMBAD database, operated at CDS, Strasbourg, France. DJM is supported in part 
by the NASA Space Grant program. The authors acknowledge the availability of 
the data made by the European Southern Observatory and Observatoire de Haute 
Provence.

\end{document}